\documentclass[12pt]{article}
\usepackage{amsmath}
\usepackage{amssymb}
\usepackage{graphicx}
\usepackage{enumerate}
\usepackage{slashed, bm}
\usepackage{hyperref, xcolor}
\definecolor{dark}{rgb}{0.10,0.2,0.3}
\definecolor{magenta}{rgb}{0.7,0.1,0.3}
\definecolor{purpure}{rgb}{0.5,0.15,0.3}
\usepackage[font=small,format=plain,labelfont=bf,up]{caption}
\usepackage{hyperref, cite}
\hypersetup{colorlinks,
citecolor=blue,
filecolor=blue,
linkcolor=magenta,
urlcolor=purpure,hyperfootnotes=true,pdftex}

\usepackage[normalem]{ulem}
\usepackage{subfig}
\usepackage{hepnames,hepunits}

\def\pt{\ensuremath{p_{\rm T}}}
\def\kt{\ensuremath{k_t}}
\def\ktz{\ensuremath{k_{t,0}}}
\newcommand{\Pmax}{\mu^2}

\usepackage{placeins}

\newcommand{\beq}{\begin{equation}}
\newcommand{\eeq}{\end{equation}}
\newcommand{\bea}{\begin{eqnarray}}
\newcommand{\eea}{\end{eqnarray}}

\newcommand{\PBproton}{PB-NLO\_ptoPb208}
\newcommand{\PBCTEQLead}{{PB-nCTEQ15FullNuc\_208\_82}}
\newcommand{\PBEPPSLead}{PB-EPPS16nlo\_CT14nl\_Pb208}

\title{%
 \vspace{-4.0cm}
 \begin{flushright}
{\small DESY 19-086,\; IFJPAN-IV-2019-7}
 \end{flushright}
 \vspace{1.5cm}
\boldmath \large \bf \PZ\ boson production in proton-lead collisions at the LHC accounting for transverse momenta of initial partons
}

\author{E.~Blanco${}^a$, A.~van~Hameren${}^a$, H.~Jung${}^b$,\\
        A.~Kusina${}^{a}$, K.~Kutak${}^{a}$
\bigskip \\
${}^a$ Institute of Nuclear Physics, Polish Academy of Sciences, \\ ul.~Radzikowskiego 152, 31-342, Cracow,
Poland\\
${}^b$ DESY, Hamburg, Germany
}

\begin{document}

\maketitle  \flushbottom

\begin{abstract}
{We perform a calculation of inclusive \PZ\ boson production in proton-lead collisions at the
LHC taking into account the transverse momenta of the initial partons. We use the framework
of $k_T$-factorization combining transverse momentum dependent parton distributions (TMDs)
with off-shell matrix elements. In order to do it we need to construct appropriate TMDs for
lead nuclei which is done using the parton branching method. Our computations are compared
with data from CMS taken at $\sqrt{s}=5.02$ TeV. The results are in good agreement with the
measurements especially the transverse momentum distribution of the \PZ\ boson.
}
\end{abstract}

\newpage
{
  \hypersetup{linkcolor=black}
 \tableofcontents
}


\section{Introduction}
\label{sec:intro}
The production of \PZ\ bosons in hadron-hadron collisions is described in lowest order (LO) calculations as the annihilation of 
a pair $\Pq\Paq \to \PZ$. In collinear factorization, the initial quarks do not carry any intrinsic transverse momentum, and therefore
the \pt\ of the \PZ\ boson vanishes. When higher order corrections in perturbative quantum chromodynamics (pQCD) are 
included, the \PZ\ boson receives a significant transverse momentum corresponding to the additional emission. However, in collinear factorization the transverse momentum 
spectrum of the \PZ\ boson at $\Lambda_{\rm QCD} < \pt < {\cal O}(10)  \GeV$ cannot be well described by a fixed order calculation, and
a resummation to all orders of soft gluon radiation is needed. Several calculations for this soft gluon resummation exist for
$pp$ and $p\bar{p}$
collisions \cite{Bertone:2019nxa,Bacchetta:2019tcu,Bizon:2018foh,Collins:1984kg}.

In a different approach of $k_T$-factorization, originally developed for small $x$ physics \cite{Catani:1990xk,Collins:1991ty}, the parton densities depend in addition on the partons'
transverse momenta. Such transverse momenta come from the intrinsic motion of the partons inside the hadrons but also from the perturbative evolution
of the partons from a small scale to the hard scale of the process. The hard process is in general calculated with off-shell initial partons. 
In the past years, significant progress has been made by the calculations of hard processes not only for initial gluons but also for initial quarks~\cite{vanHameren:2016kkz}.
The transverse momentum dependent parton densities (TMDs) for protons were recently obtained from precision fits to deep-inelastic cross section 
measurements within the parton-branching (PB) approach \cite{Hautmann:2017fcj,Hautmann:2017xtx,Martinez:2018jxt}.

In this paper we are in particular interested in exploring the transverse momentum structure of the partonic content of lead nucleus at relatively large values of its longitudinal momentum.\footnote{This allows us to work with linear evolution equations and not to be affected by the saturation of gluon densities \cite{Gribov:1984tu}.}
To achieve this we extend the PB approach to the case of heavy nuclei, in particular to lead nucleus, and apply the newly constructed nuclear TMDs (nTMDs)
together with off-shell matrix elements to calculations of $\PZ$ boson production in $p$Pb  collisions at the LHC.%
    \footnote{Whenever we refer to $\PZ$ boson production we mean a production
    of a lepton pair via both $\PZ$ and $\gamma^*$ exchange.}

The interest is motivated by experiments at CERN where proton-lead and lead-lead collisions are studied. The precise knowledge of the partonic structure of the lead nucleus and the factorization used will allow to increase the precision of the theoretical description of the initial state of proton-lead and lead-lead collisions. 
The interest in knowing the precise transverse momentum structure of the nucleus lies in the following:
when the quark gluon plasma is produced in lead-lead collision the propagating jet broadens and gets kicks from the plasma constituents. Therefore, its transverse spectrum changes. Moreover the initial state effects (initial state shower) lead to decorrelations. The precise knowledge of the transverse momentum distribution of partons in lead nucleus will therefore be beneficial for a more precise determination of final-state effects due to jet-plasma interactions. 
In order to demonstrate the usefulness of the newly obtained nTMDs for lead nucleus, we calculate the cross section for the rapidity and $p_T$ spectrum of Drell-Yan pairs with an intermediate $\PZ/\gamma^*$ boson state.
Furthermore, such a final state, being a colorless particle, gives the opportunity for particularly interesting investigations complementary to results obtained in studies of
jet final states in \cite{Bury:2017jxo,Deak:2018obv}.

\section{Nuclear TMDs}
\label{sec:pdfs}
In the PB approach, the parton density is evolved with the DGLAP evolution equation from a small scale (where the 
initial parton density is parametrized) to the scale of the hard process using an iterative procedure. In this way, every single
splitting process during the evolution is calculated, and kinematic constraints in each parton splitting step 
are treated. Once a physical meaning is given to the evolution scale, the transverse momentum of the partons involved in
each splitting can be calculated, and a transverse momentum dependent (TMD) parton density can be obtained.

We apply DGLAP splitting functions at next-to-leading order (NLO) and we use angular ordering for the parton evolution, which 
relates the evolution scale $\mu_i$ to the transverse momentum of the emitted parton  ${\bf q}_{t,i}^2  =  (1-z_i)^2 \mu_i^ 2$, where
$z_i$ is the fractional momentum in the splitting process (details of the procedure are described in Ref. \cite{Martinez:2018jxt}).

The nuclear TMD, ${\cal A}^{\mathrm{Pb}}_a(x,\kt^2,\mu^2)$, is obtained by a convolution
of the starting distribution ${\cal A}^{\mathrm{Pb}}_{0,b} (x',\ktz^2,\mu_0^2)$ with the
evolution kernel ${\cal K}_{b a}\left(x'',\ktz^2,\kt^2,\mu_0^2,\Pmax\right)$ as described
in Ref.~\cite{Martinez:2018jxt} (with $k_t^2 = {{\bf k }^2 }$):
\begin{multline}
x{\cal A}^{\mathrm{Pb}}_a(x,\kt^2,\mu^2) 
 = x\int dx' \int dx'' {\cal A}^{\mathrm{Pb}}_{0,b} (x',\ktz^2,\mu_0^2) {\cal  K}_{ba}\left(x'',\ktz^2,\kt^2,\mu_0^2,\Pmax\right) 
 \delta(x' x'' - x) 
\\
 =  \int dx' {\cal A}^{\mathrm{Pb}}_{0,b} (x',\ktz^2,\mu_0^2)
\frac{x}{x'} \ { {\cal  K}_{ba}\left(\frac{x}{x'},\ktz^2,\kt^2,\mu_0^2,\Pmax\right) }  \;\; .
\label{TMD_kernel}
\end{multline}
The evolution  of the kernel starts at $x_0=1$ at $\mu_0^2$. 
In general, the starting distribution ${\cal A}_0$ can have flavor and $x$ dependent $\ktz$ distributions, for simplicity we use here 
a factorized form:
\begin{equation}
{\cal A}^{\mathrm{Pb}}_{0,b} (x,\ktz^2,\mu_0^2) = f^{\mathrm{Pb}}_{0,b} (x,\mu_0^2) \cdot \exp(-| \ktz^2 | / \sigma^2)
\label{TMD_A0}
\end{equation}
where the intrinsic $\ktz$ distribution is given by a Gauss distribution
with $ \sigma^2  =  q_0^2 / 2 $  for all flavors and all $x$ with a constant
value $q_0 = 0.5$~\GeV. 
The $q_0$ value is assumed to be the same as in the proton TMD
fit~\cite{Martinez:2018jxt}.

The starting distribution $f^{\mathrm{Pb}}_{0,b} (x,\mu_0^2)$ for lead nucleus
is taken to be one of the available collinear nuclear PDFs (nPDFs),
e.g.\ the nCTEQ15~\cite{Kovarik:2015cma}.
We always produce two sets of nTMDs which differ by the initial scale $\mu_0^2$
and the argument of the strong coupling, $\alpha_s$. We refer to these sets as
{\tt Set1} and {\tt Set2}, where {\tt Set1} features $\mu_0^2=1.9$ GeV$^2$ and
the scale of $\alpha_s$ coinciding with the DGLAP evolution variable in~\cite{Martinez:2018jxt}.
On the other hand {\tt Set2} has $\mu_0^2=1.4$ GeV$^2$ and the $\alpha_s$ scale
is chosen to be the transverse momentum $|{\bf q}^2_{t,i}|$ of the splitting
process, as suggested in Refs.~\cite{Amati:1980ch,Gieseke:2003rz,Ciafaloni:2003rd}
and also applied in~\cite{Martinez:2018jxt}.
This choice is motivated by accounting for coherence effects effectively introducing higher order corrections.
For the purpose of practical calculations in Sec.~\ref{sec:results} we used
three different collinear nPDFs to produce the corresponding nTMDs. We adopted
the two most commonly used nPDFs nCTEQ15~\cite{Kovarik:2015cma} and EPPS16~\cite{Eskola:2016oht}.
Additionally we did comparisons with distributions obtained in Ref.~\cite{Kusina:2017gkz}
which allow for more reliable description of the low-$x$ region.
However, the nTMDs based on these nPDFs (PB-gluon\_D\_c\_ncteq1568CL\_Pb)
turned out to be rather similar to the results obtained based on the nCTEQ15 nPDFs
which is why we show them only later when discussing results for \PZ\ boson production.

\begin{figure}[!ht]
\begin{center}
\subfloat[]{
\includegraphics[width=0.5\textwidth]{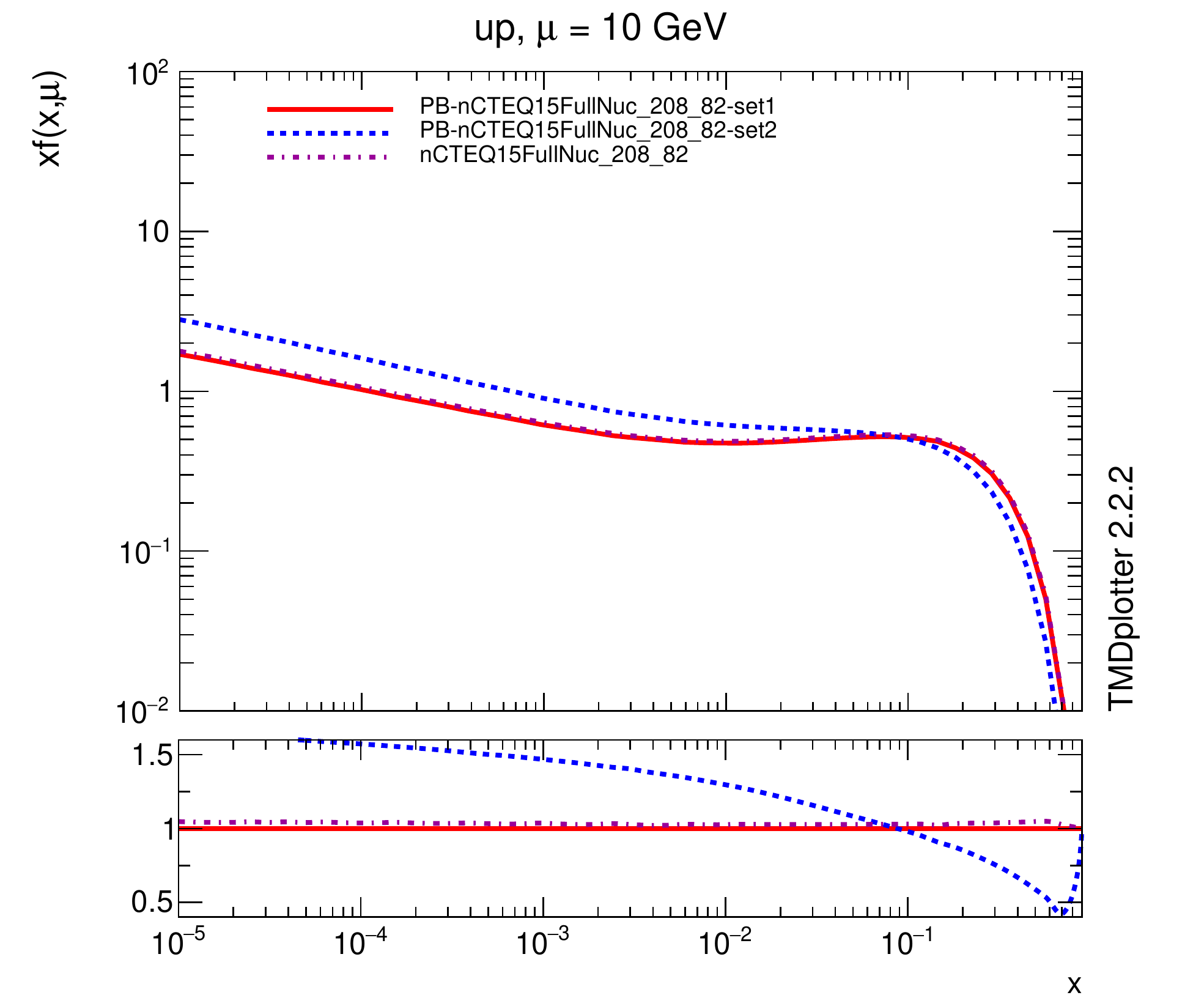}
\label{fig:collinear_pdfs_check10}}
\subfloat[]{
\includegraphics[width=0.5\textwidth]{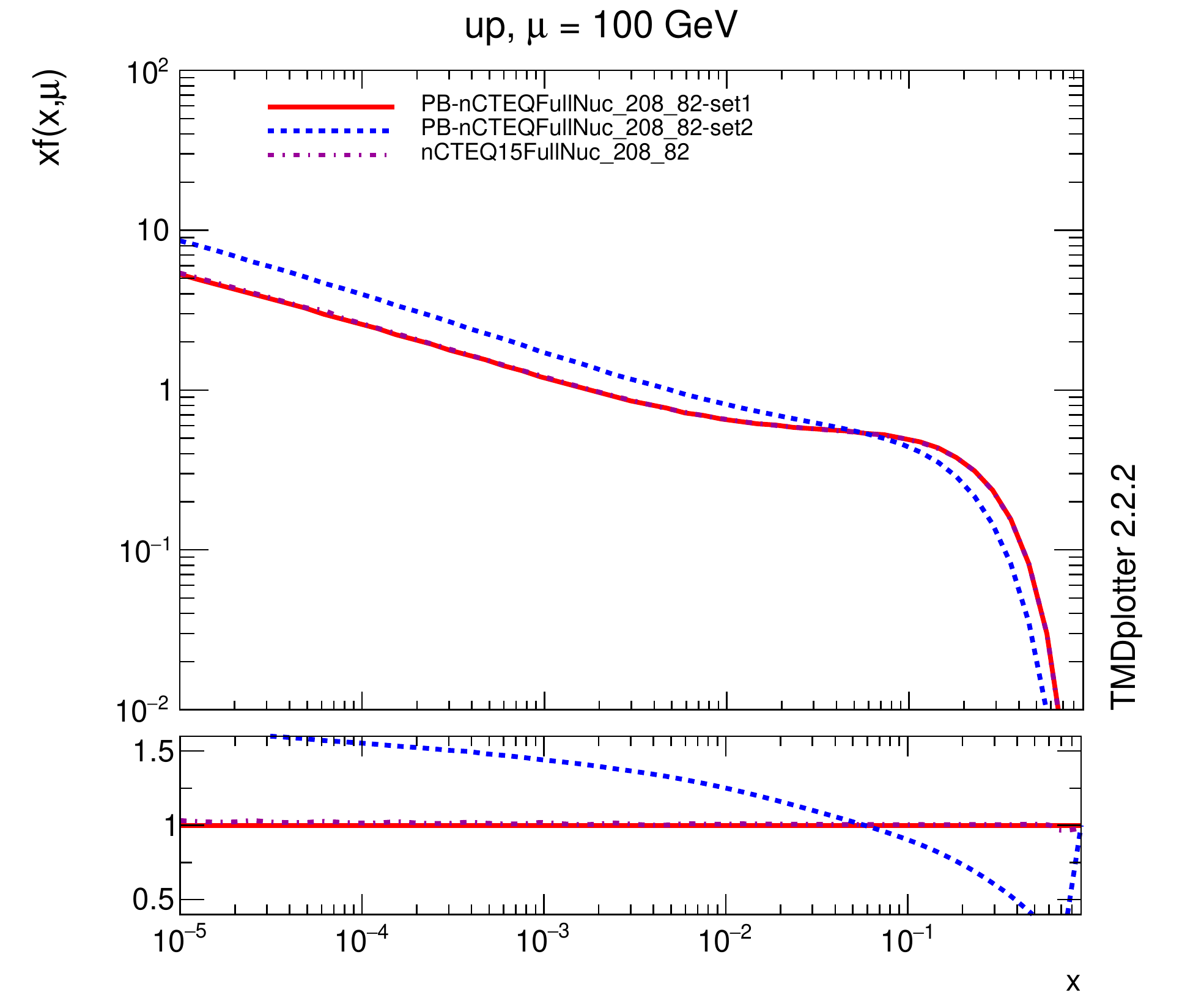}
\label{fig:collinear_pdfs_check100}}
\\
\subfloat[]{
\includegraphics[width=0.5\textwidth]{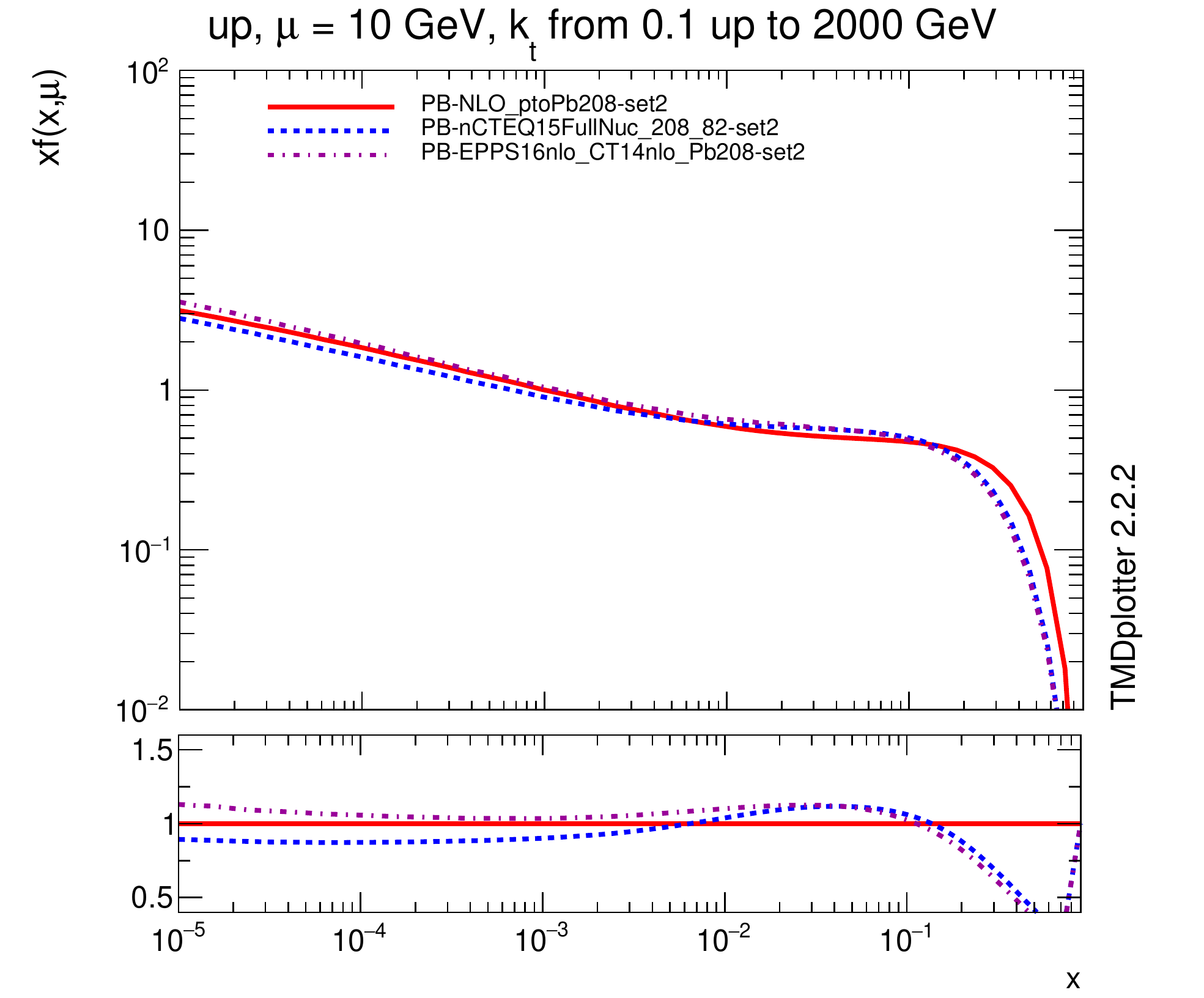}
\label{fig:collinear_pdfs_compare10}}
\subfloat[]{
\includegraphics[width=0.5\textwidth]{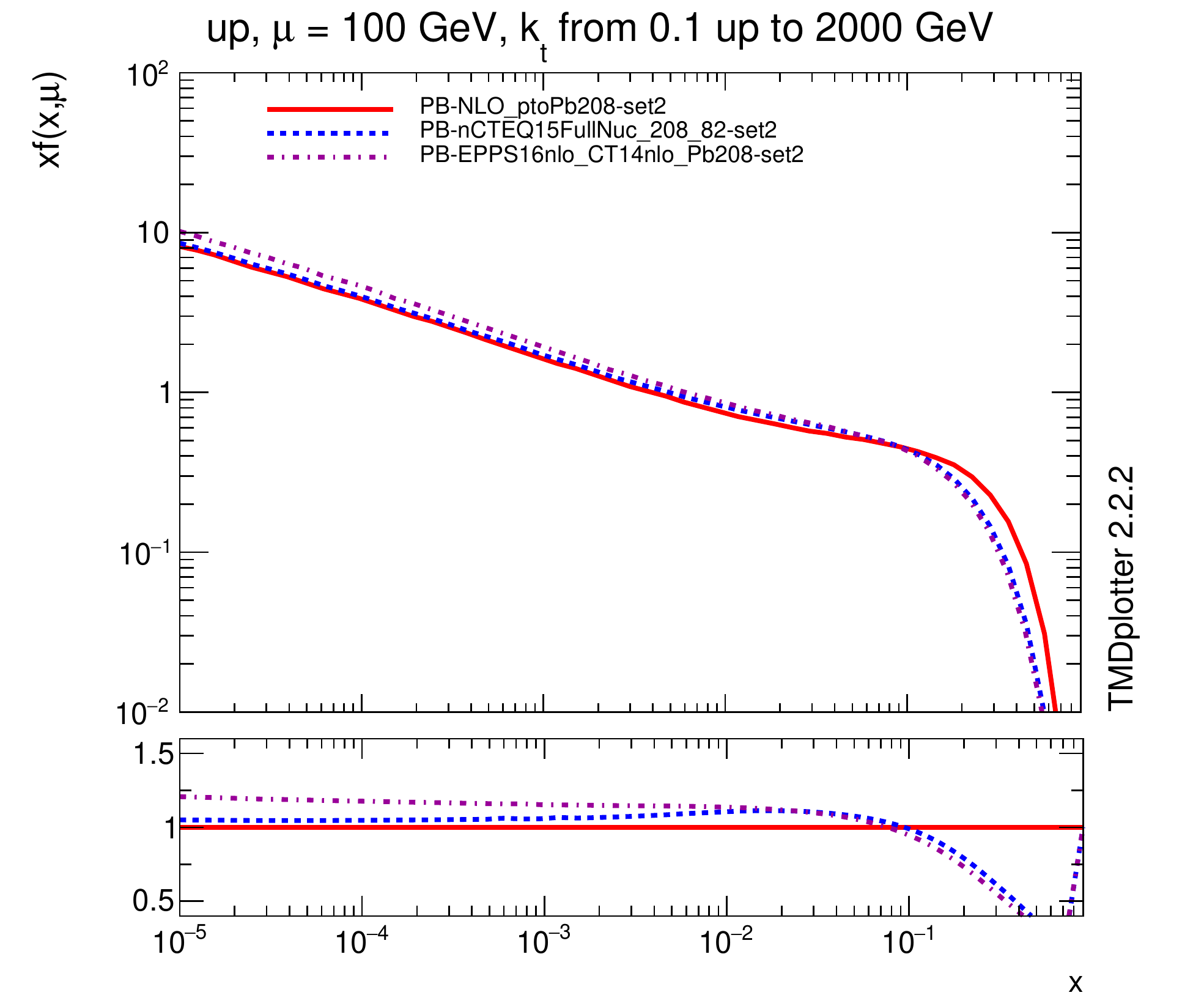}
\label{fig:collinear_pdfs_compare100}}
\caption{\small Up quark transverse momentum dependent parton density integrated
over \kt\ for different values of the scale $\mu$. The upper plots show the
comparison with the corresponding collinear distribution. The lower plots compare
two nuclear TMD distributions based on the nCTEQ15 (\PBCTEQLead ) and EPPS16
(\PBEPPSLead ) nuclear PDFs with the \PBproton  TMD.
}
\label{fig:collinear_pdfs}
\end{center}
\end{figure}
We now present the obtained  nTMDs.
In Fig.~\ref{fig:collinear_pdfs} we show the \kt\ integrated distributions of the up quark
at the scale $\mu=10$ GeV and $\mu=100$ GeV as a function of $x$.
First in Figs.~\ref{fig:collinear_pdfs_check10} and~\ref{fig:collinear_pdfs_check100}
we plot the distributions obtained form the nCTEQ15 starting distribution evolved with the PB method  using different scales
in $\alpha_S$ according to the {\tt Set1} and {\tt Set2} prescriptions (\PBCTEQLead ), and compare them
with the original nCTEQ15 collinear PDFs. One can observe clear differences between the
{\tt Set1} and {\tt Set2} distributions, but as expected after integration over the $k_t$ the
{\tt Set1} distributions reproduce the collinear PDFs.
In the computations that will follow in Sec.~\ref{sec:results} we mostly concentrate on the {\tt Set2} distributions which are preferred by the phenomenological applications~\cite{Martinez:2018jxt}.
In Figs.~\ref{fig:collinear_pdfs_compare10} and~\ref{fig:collinear_pdfs_compare100}
we also compare the \PBCTEQLead\ {\tt Set2} distributions with another {\tt Set2}
TMDs obtained by using the EPPS16 lead nPDFs as starting distributions~\cite{Eskola:2016oht}
(\PBEPPSLead ) and with the \PBproton\ proton TMDs%
    \footnote{More precisely, in order to have a meaningful comparison with the proton TMDs,
    in this case we construct a combination of protons and neutrons distributions forming
    a lead nuclei: $f^{\mathrm{Pb}}=82/208f^p+(208-82)/208f^n$, with the neutron distribution
    obtained assuming isospin symmetry.}
obtained in Ref.~\cite{Martinez:2018jxt}, which we will also employ in our calculations.

\begin{figure}[!t]
\begin{center} 
\includegraphics[width=0.49\textwidth]{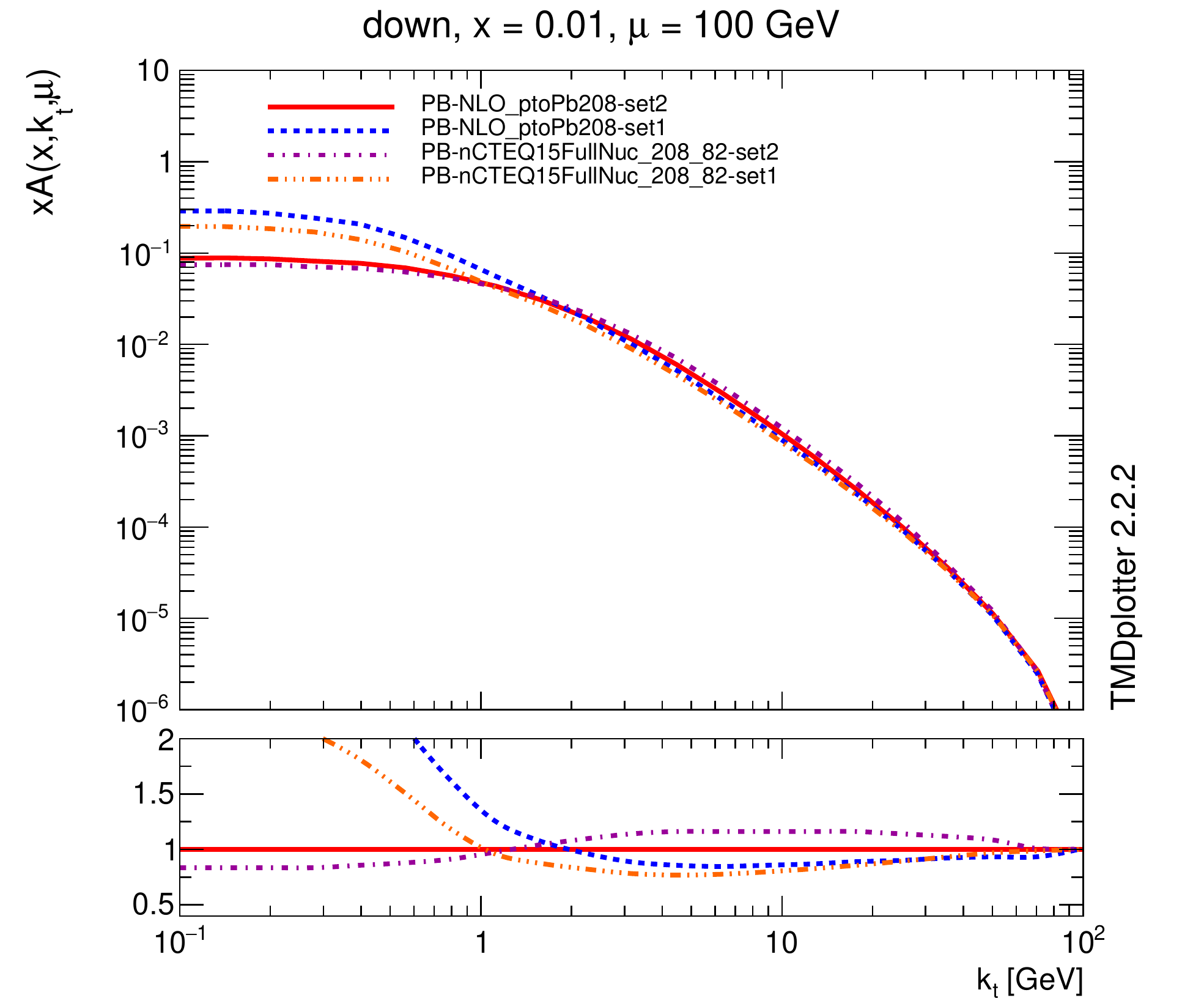}
\includegraphics[width=0.49\textwidth]{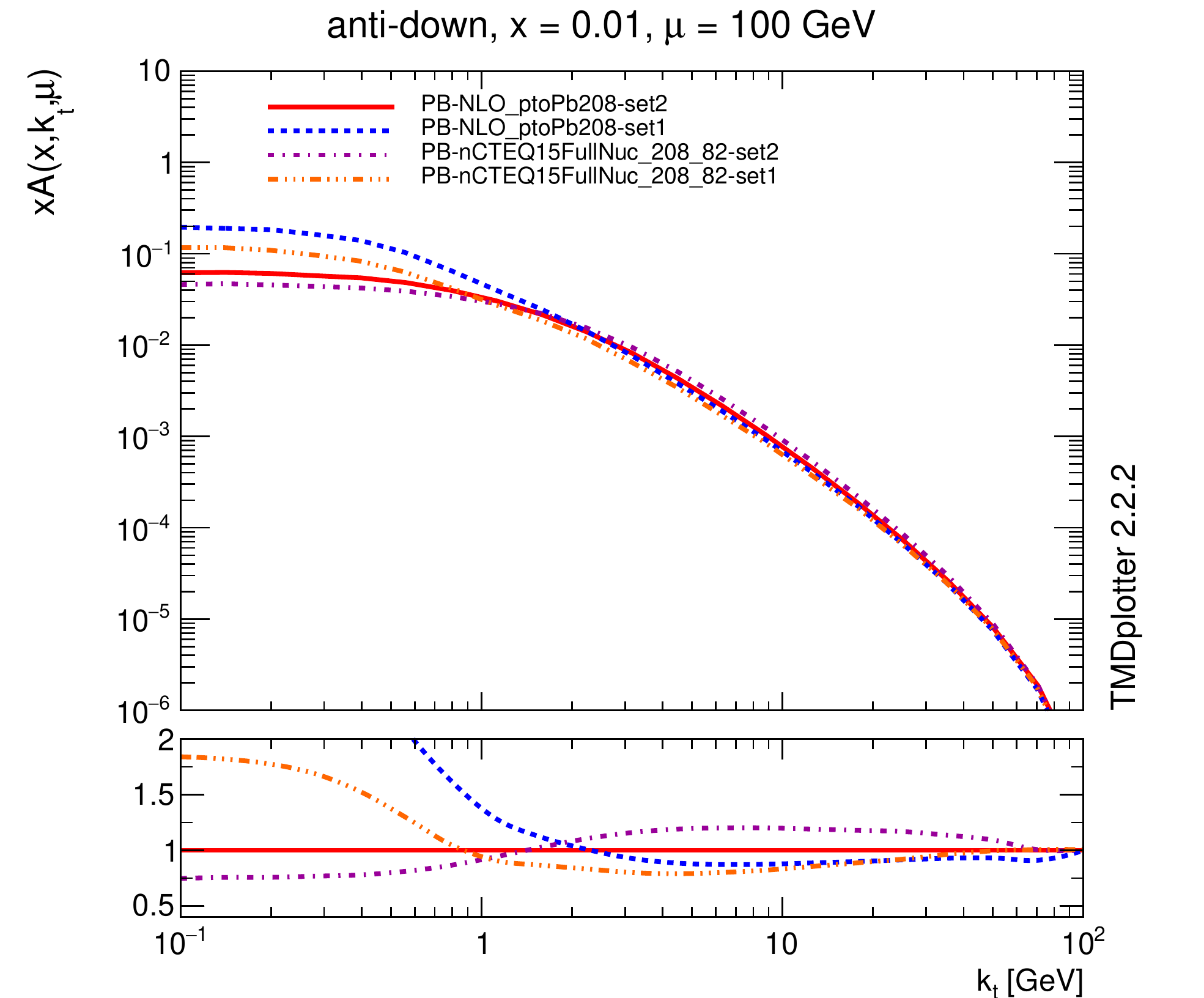}
\includegraphics[width=0.49\textwidth]{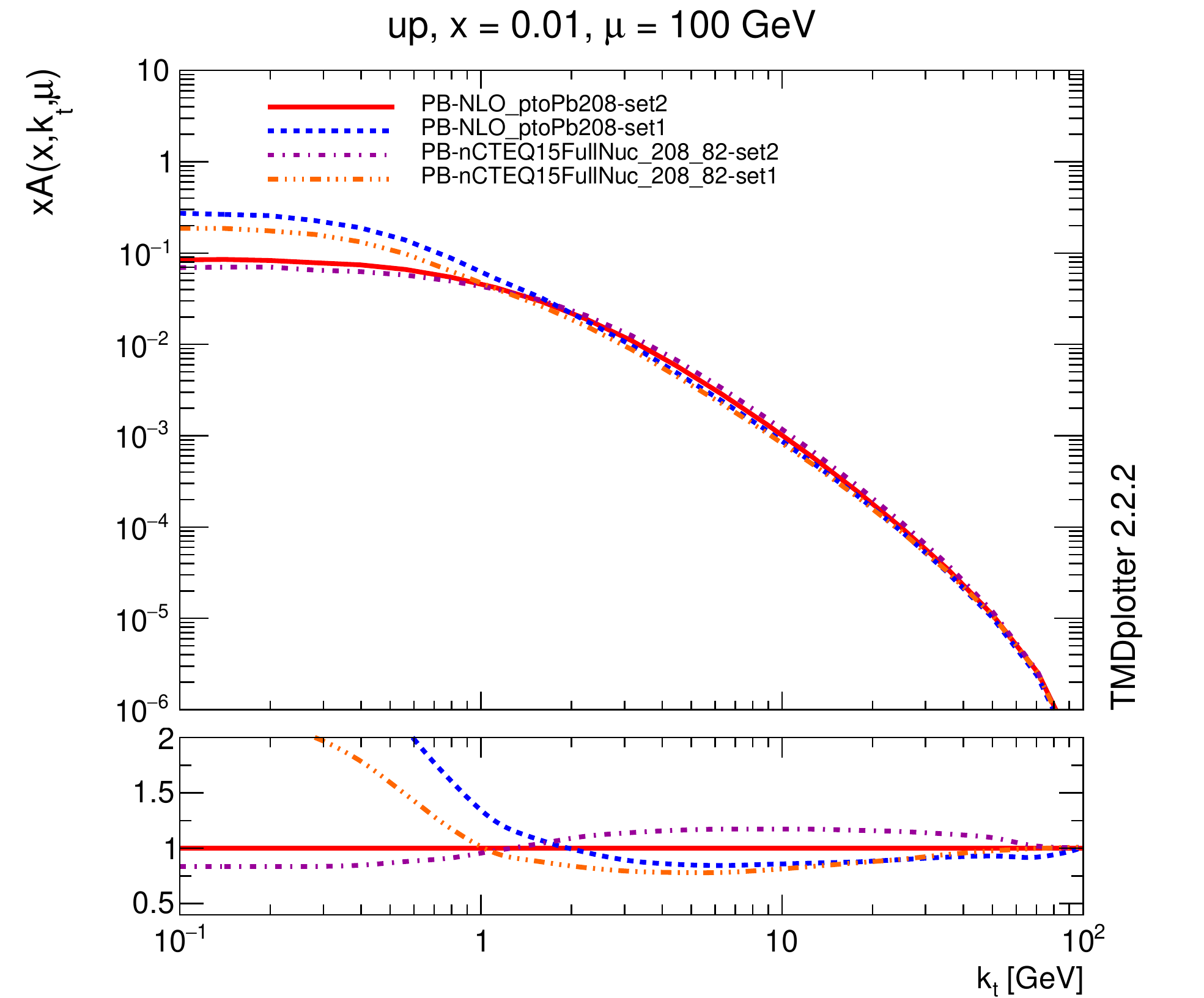}
\includegraphics[width=0.49\textwidth]{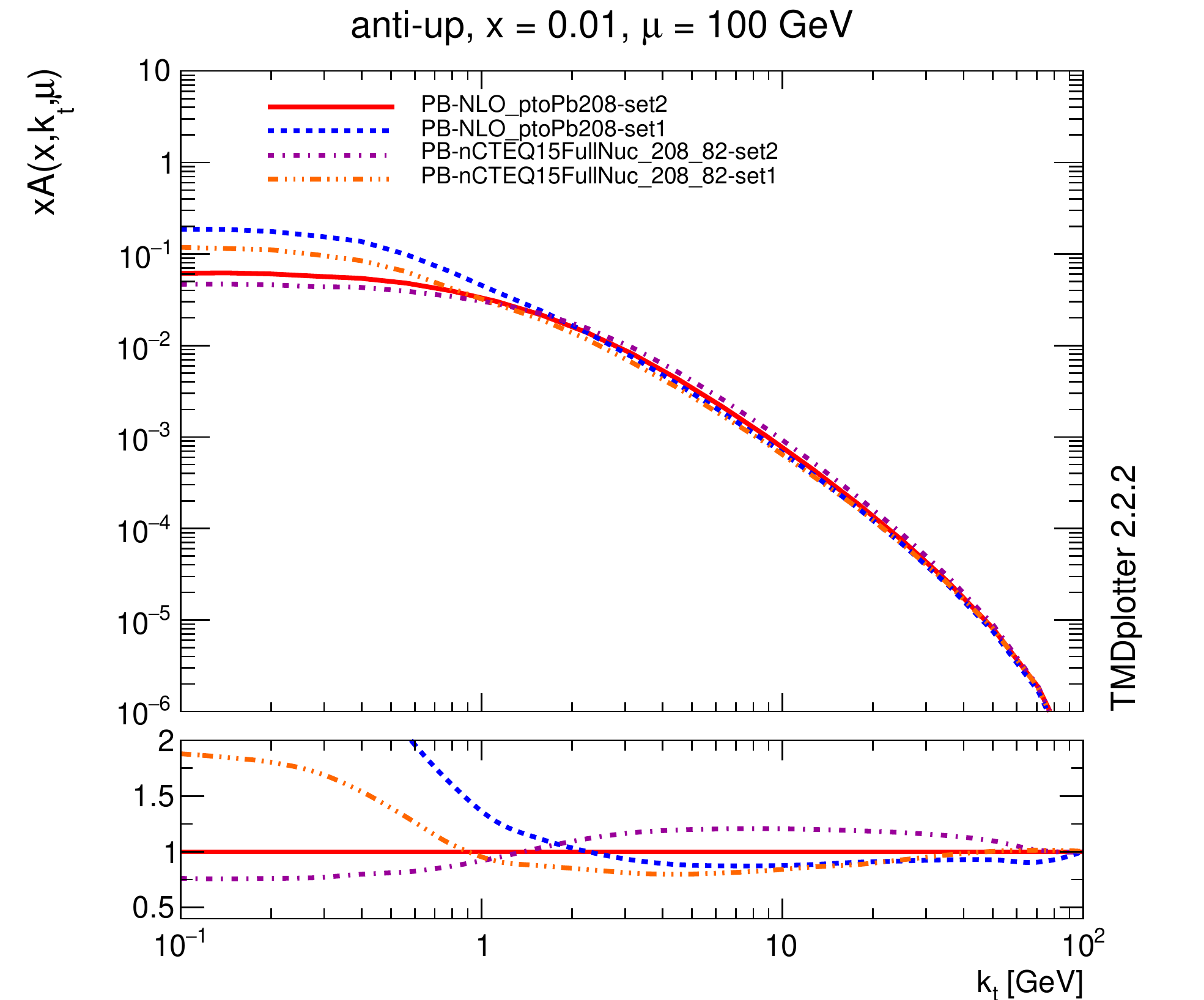}
\caption{\small Transverse momentum dependent parton densities for different
quark species at $x=0.01$ at the scale of $\mu=100$ GeV.
The ratio is always taken with respect to the \PBproton\ {\tt set2} distributions.
}
\label{TMD_pdfs}
\end{center}
\end{figure}

\begin{figure}[!t]
\begin{center} 
\includegraphics[width=0.49\textwidth]{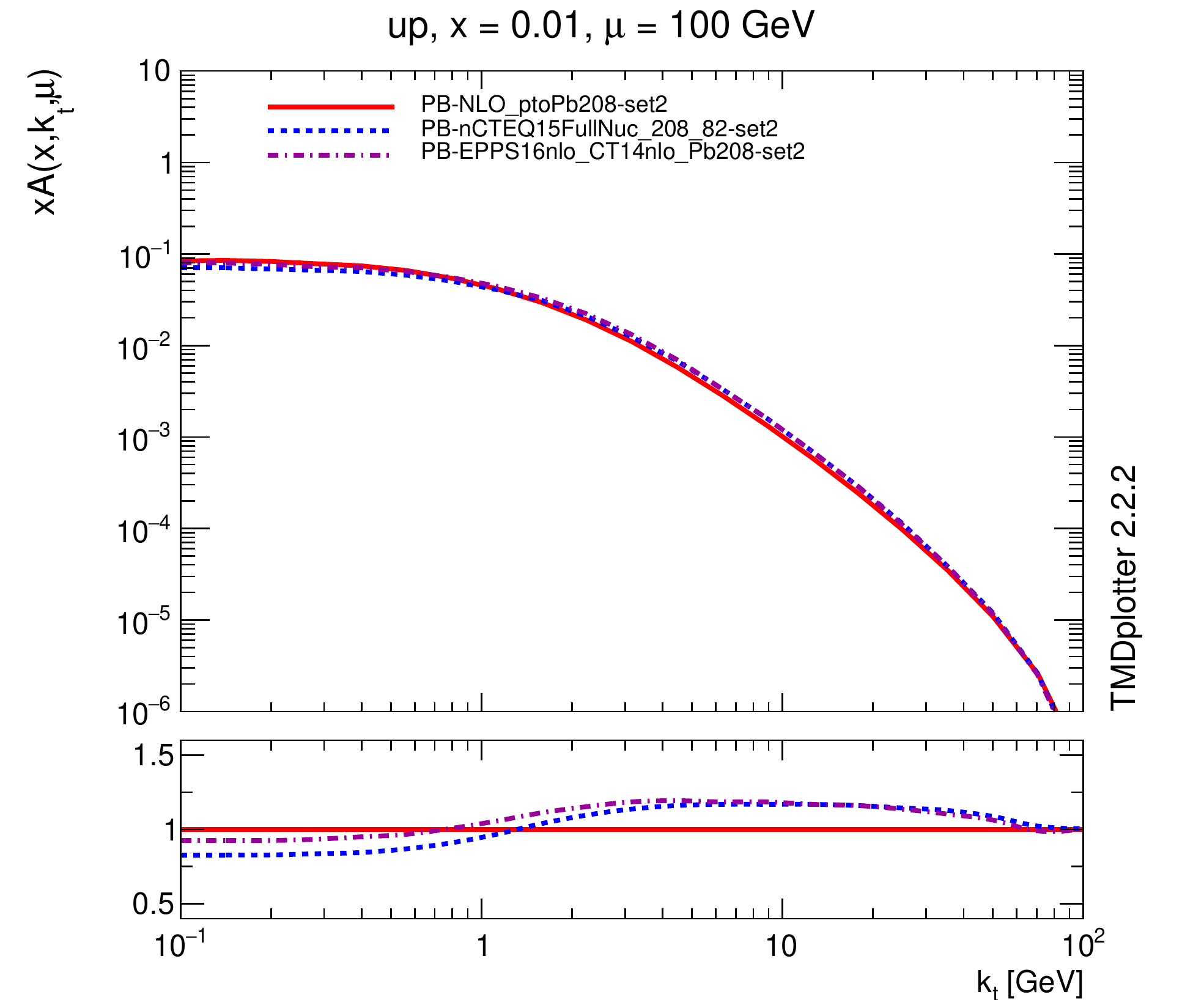}
\includegraphics[width=0.49\textwidth]{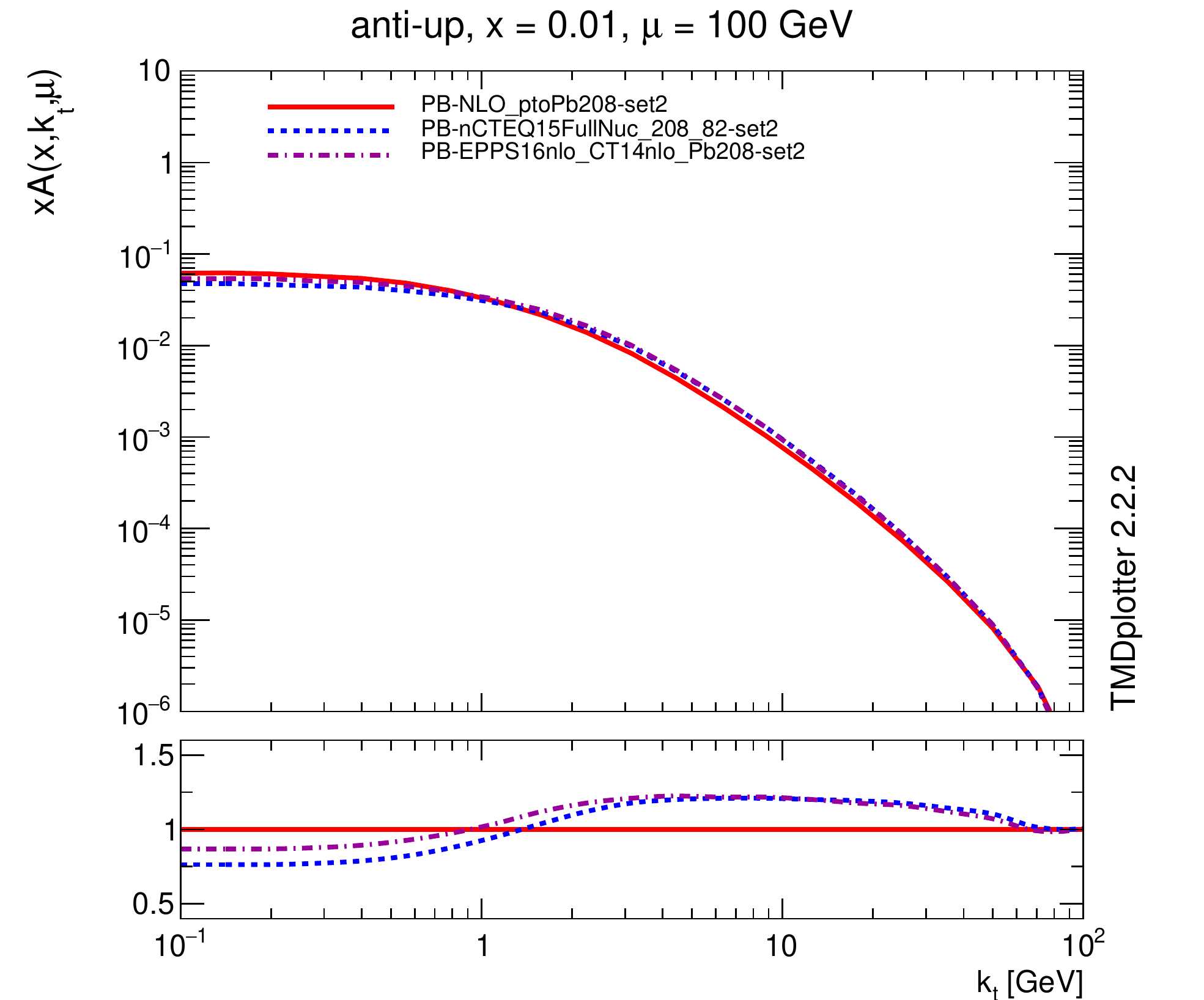}
\caption{\small Comparison of different nuclear transverse momentum dependent
parton densities for $u$ and $\bar{u}$ quarks at $x=0.01$ at the scale $\mu=100$ GeV.
Depicted are the {\tt Set2} distributions that are used in the calculations in
Sec.~\ref{sec:results}. The ratio is taken with respect to the proton \PBproton\
{\tt set2} TMD.
}
\label{TMD_pdfs_com}
\end{center}
\end{figure}

\begin{figure}[!t]
\begin{center}
\includegraphics[width=0.32\textwidth]{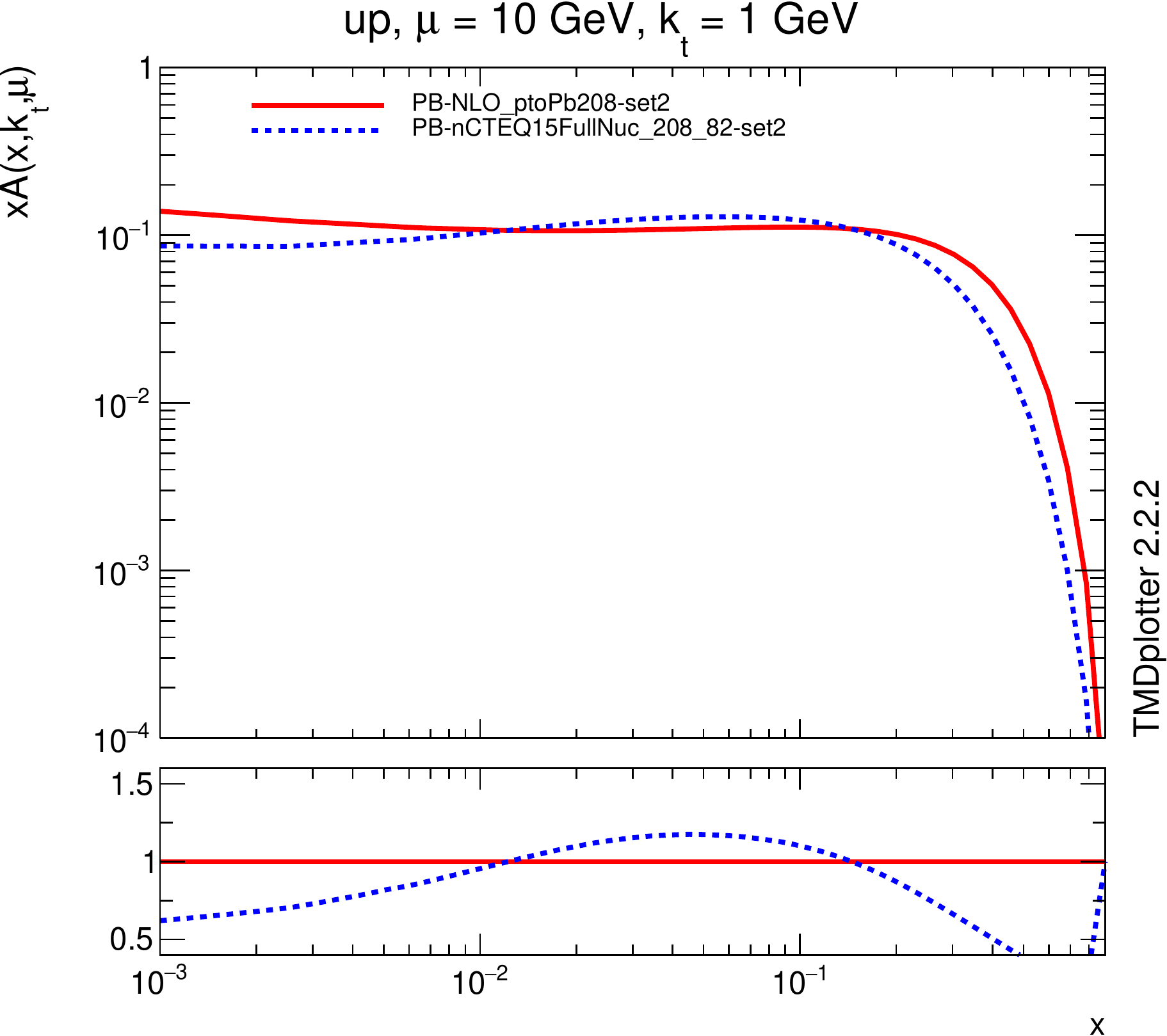}
\includegraphics[width=0.32\textwidth]{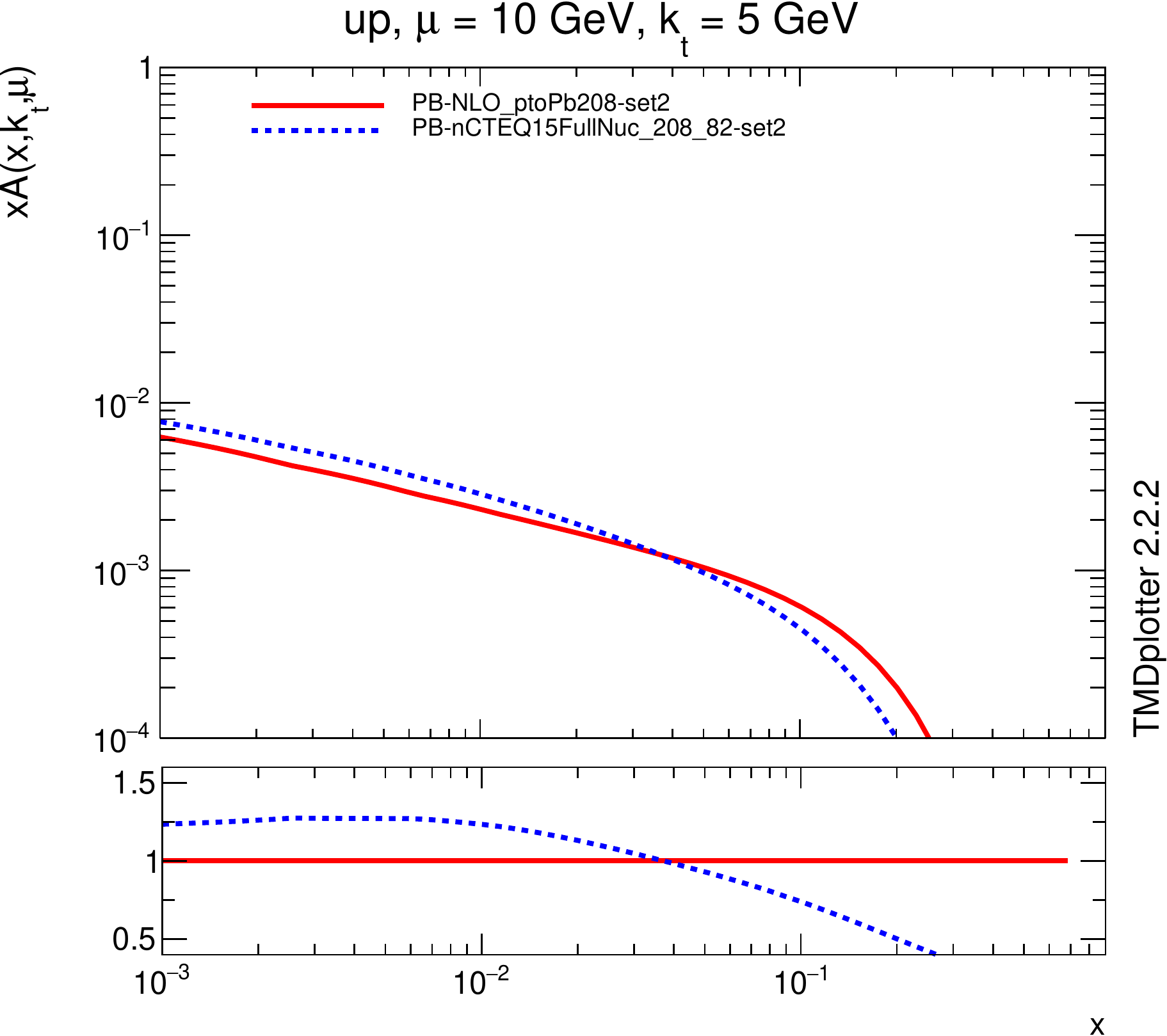}
\includegraphics[width=0.32\textwidth]{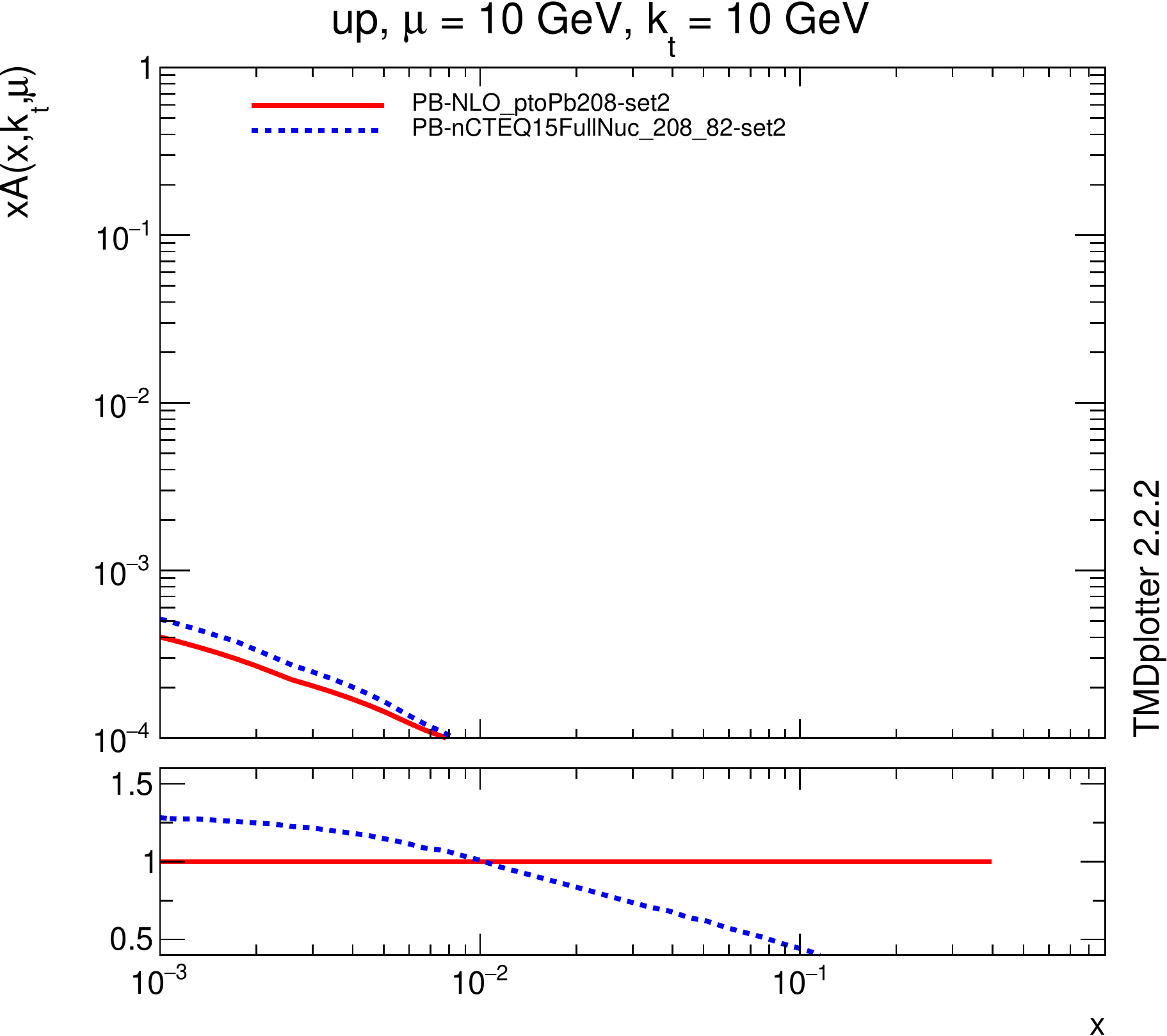}
\\
\includegraphics[width=0.32\textwidth]{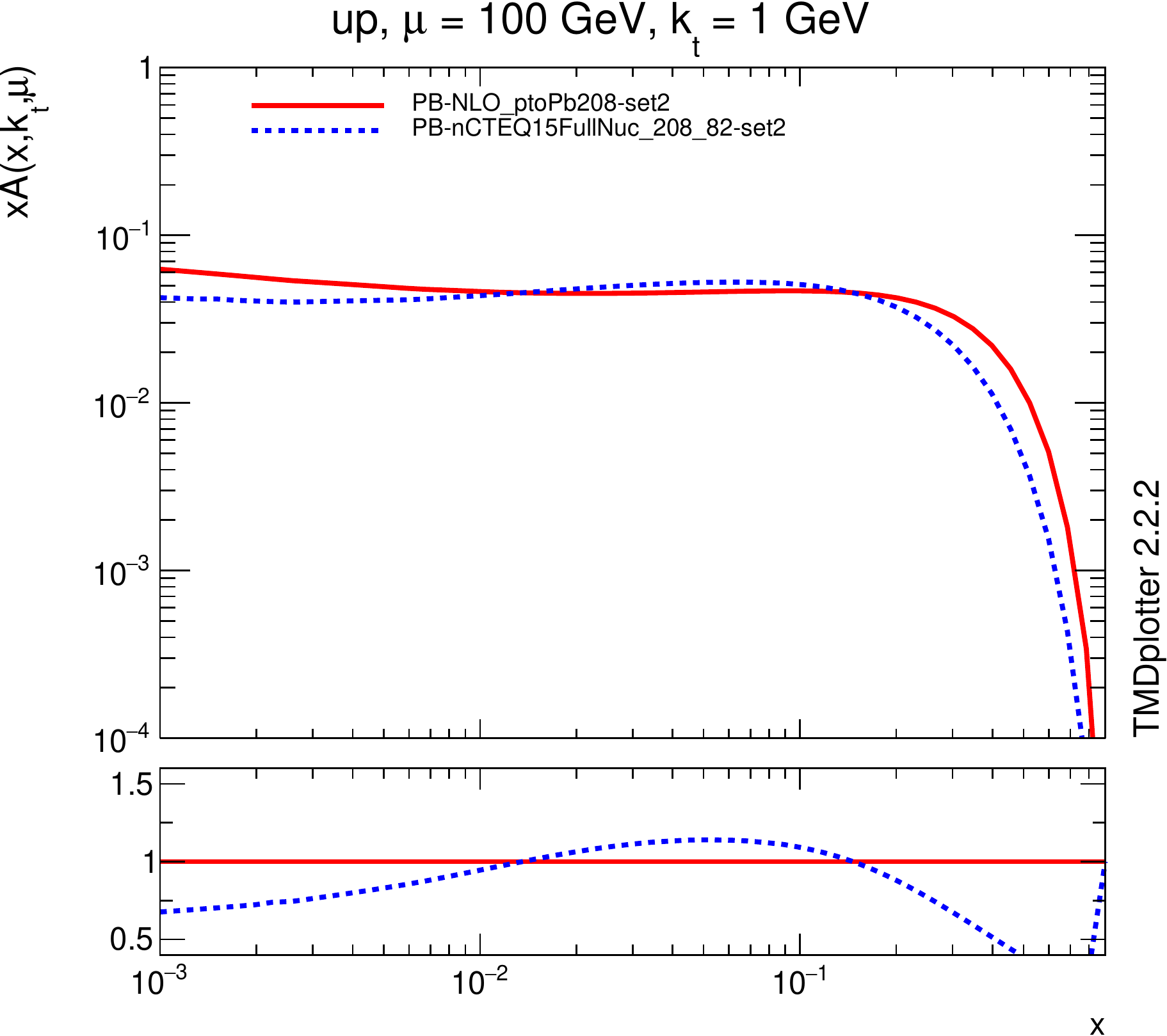}
\includegraphics[width=0.32\textwidth]{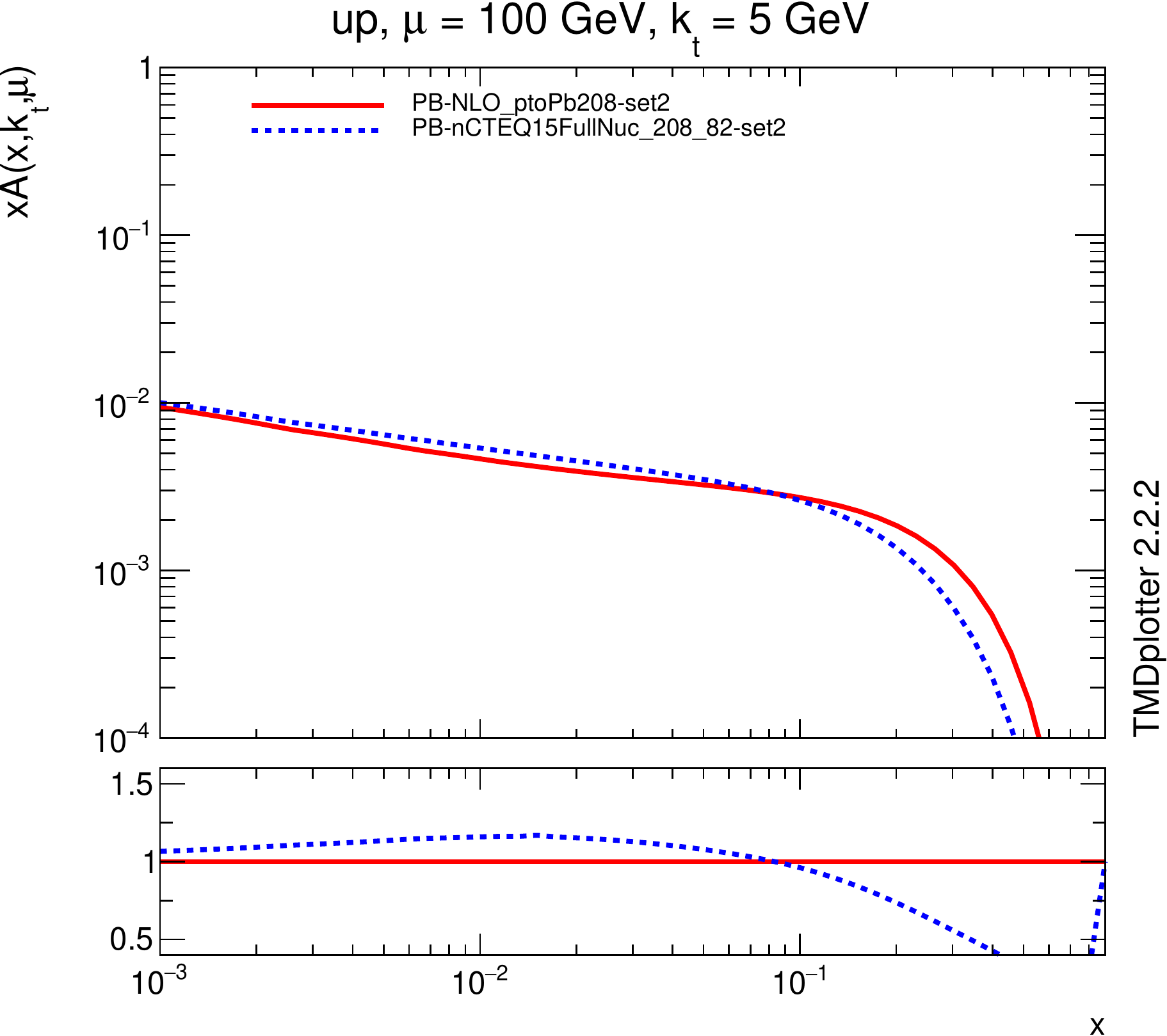}
\includegraphics[width=0.32\textwidth]{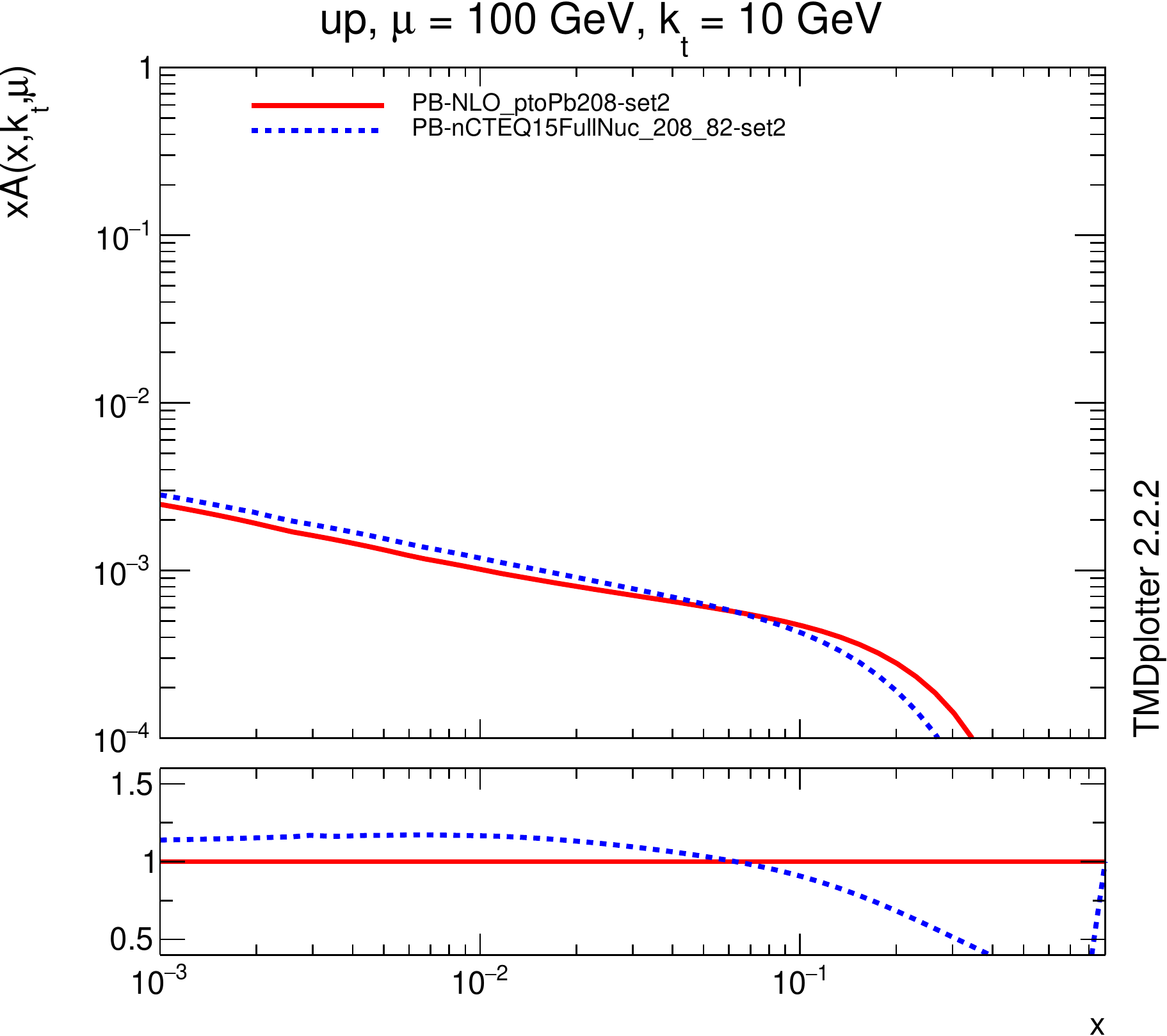}
\caption{\small
Comparison of {\tt set2} nuclear \PBCTEQLead\ and 
proton \PBproton\ TMDs for up quark at several $k_t$ and $\mu$ scales
as a function of $x$. The ratio plots take as a reference the proton
TMDs (showing the nuclear modification factors).
}
\label{TMD_pdfs_com_nucmod}
\end{center}
\end{figure}

In Fig.~\ref{TMD_pdfs} the transverse momentum dependent parton distributions for different
quark species ($u$, $d$, $\bar{u}$, $\bar{d}$) are shown at $x=0.01$ and $\mu=100$ GeV
as a function of the transverse momentum. We show here the \PBCTEQLead\  TMDs for {\tt Set1}
and {\tt Set2} and compare them with the corresponding combination of proton TMDs \PBproton .
%
%
In Fig.~\ref{TMD_pdfs_com} we compare different nuclear TMDs with   \PBproton\ 
distributions restricting only to the {\tt Set2} case at $x=0.01$ and $\mu=100$ GeV.
We can see that both nuclear TMDs (\PBCTEQLead\   and \PBEPPSLead ) are quite similar
and differ from the \PBproton\ distributions.
We can see that for small values of transverse momentum (below 1~GeV) there is
a supression of the nTMDs compared to the proton TMDs. On the other hand, when
the $k_t$ rises above 1~GeV we start to observe an enhancement of nTMDs which
presists up to the high $k_t$ values ($\sim80$ GeV). A similar behaviour is observed
for other $x$ values and other flavours.
To even further quantify the nuclear modifications of TMDs,
in Fig.~\ref{TMD_pdfs_com_nucmod} we provide additional distributions
for different $\mu$ and $k_t$ values as a function of $x$.
At lower $k_t$ values we can observe a behaviour similar to the collinear
PDFs which exhibit a rather well defined shawoding, anti-shadowing and EMC regions.
\\

\begin{figure}[!t]
\begin{center} 
\includegraphics[width=0.49\textwidth]{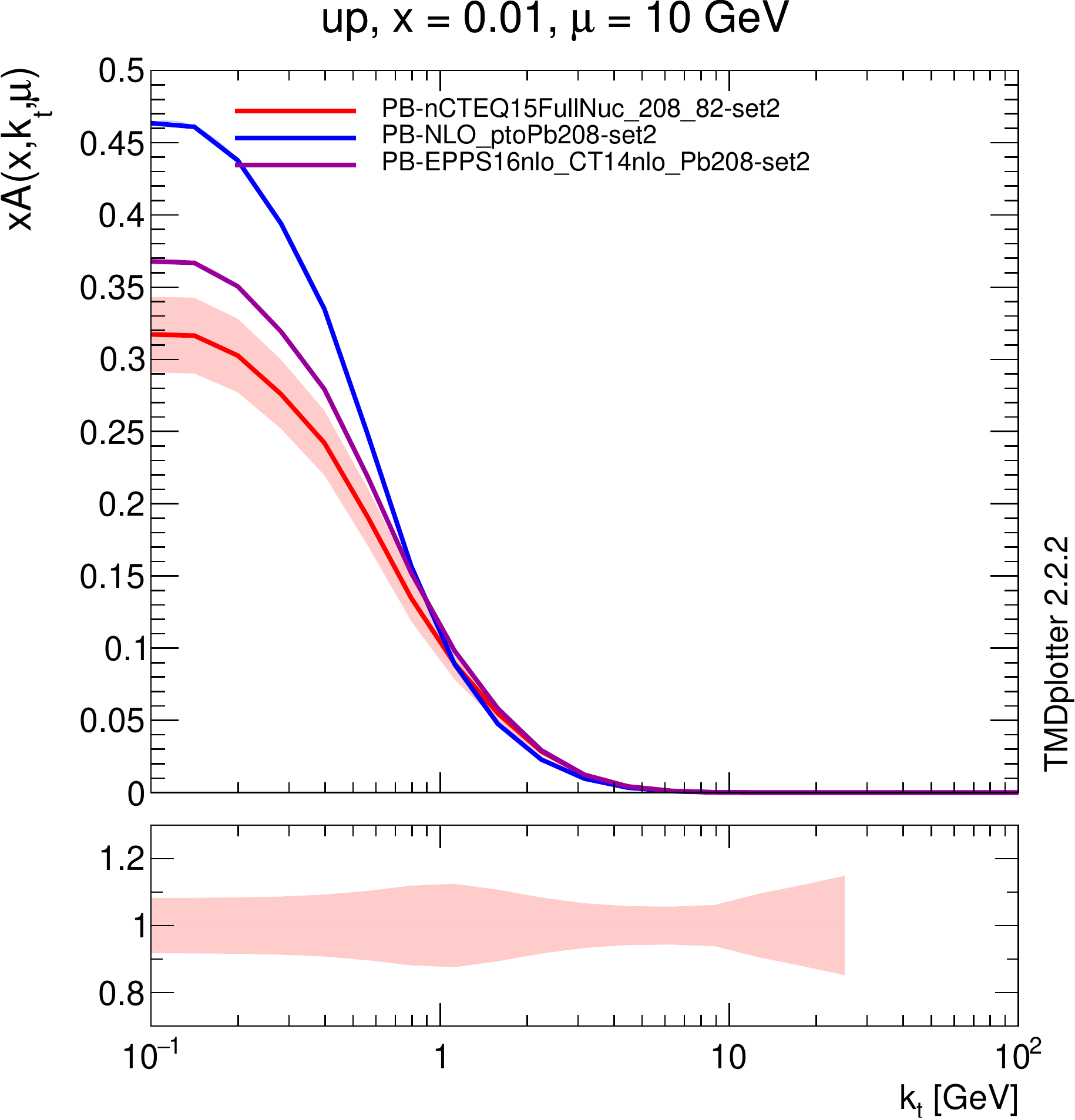}
\hfil
\includegraphics[width=0.49\textwidth]{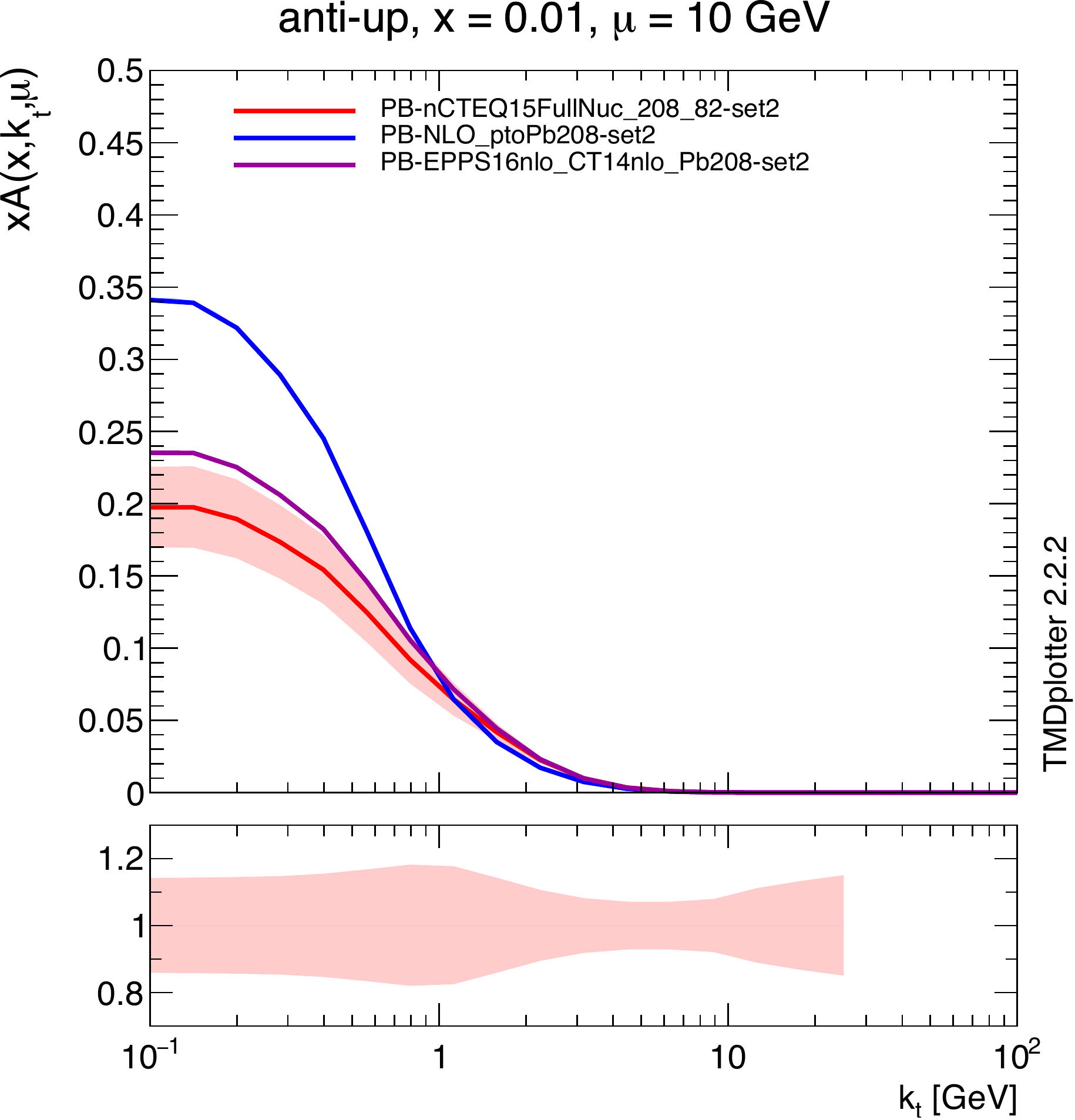}
\\
\includegraphics[width=0.49\textwidth]{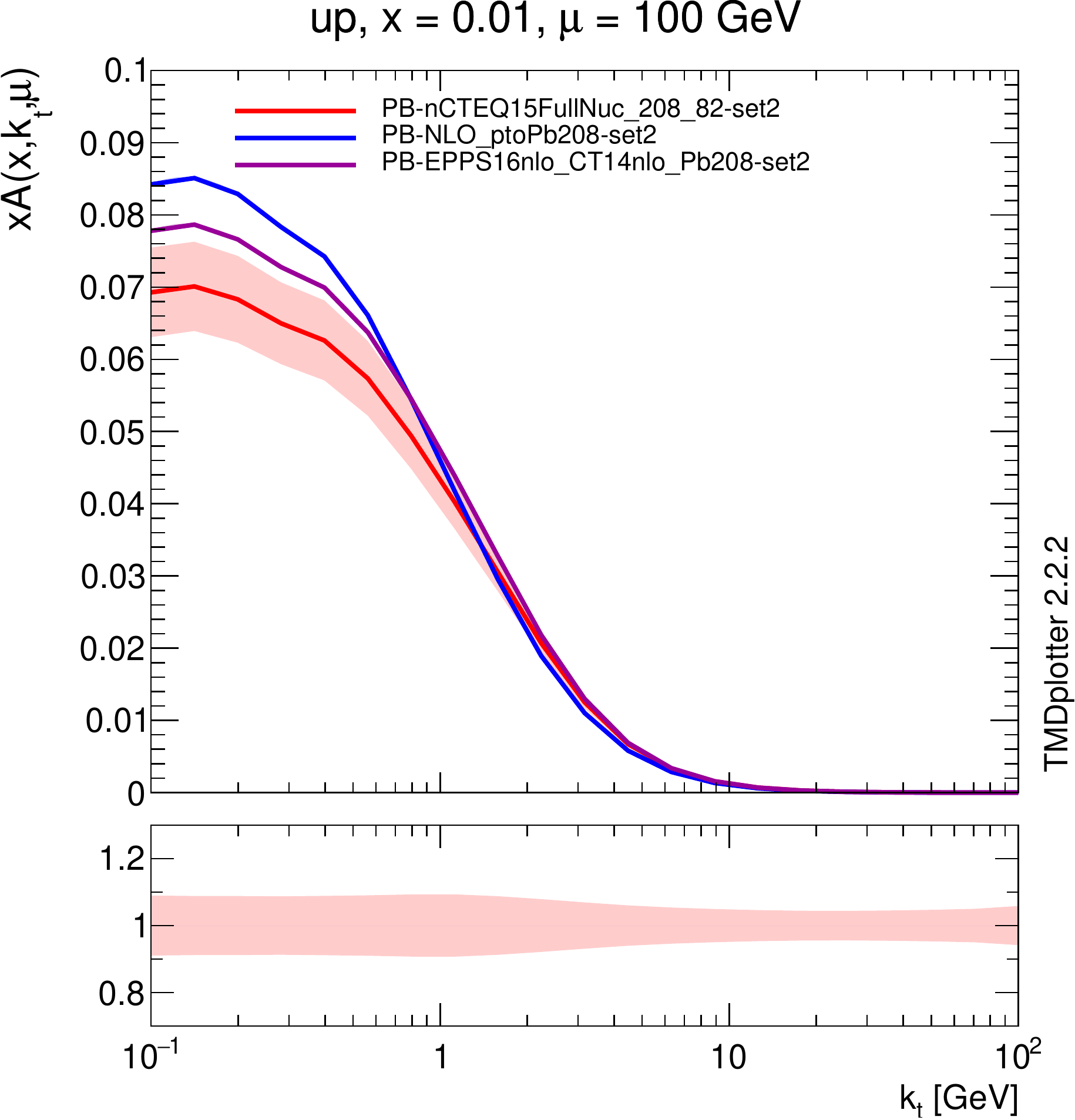}
\hfil
\includegraphics[width=0.49\textwidth]{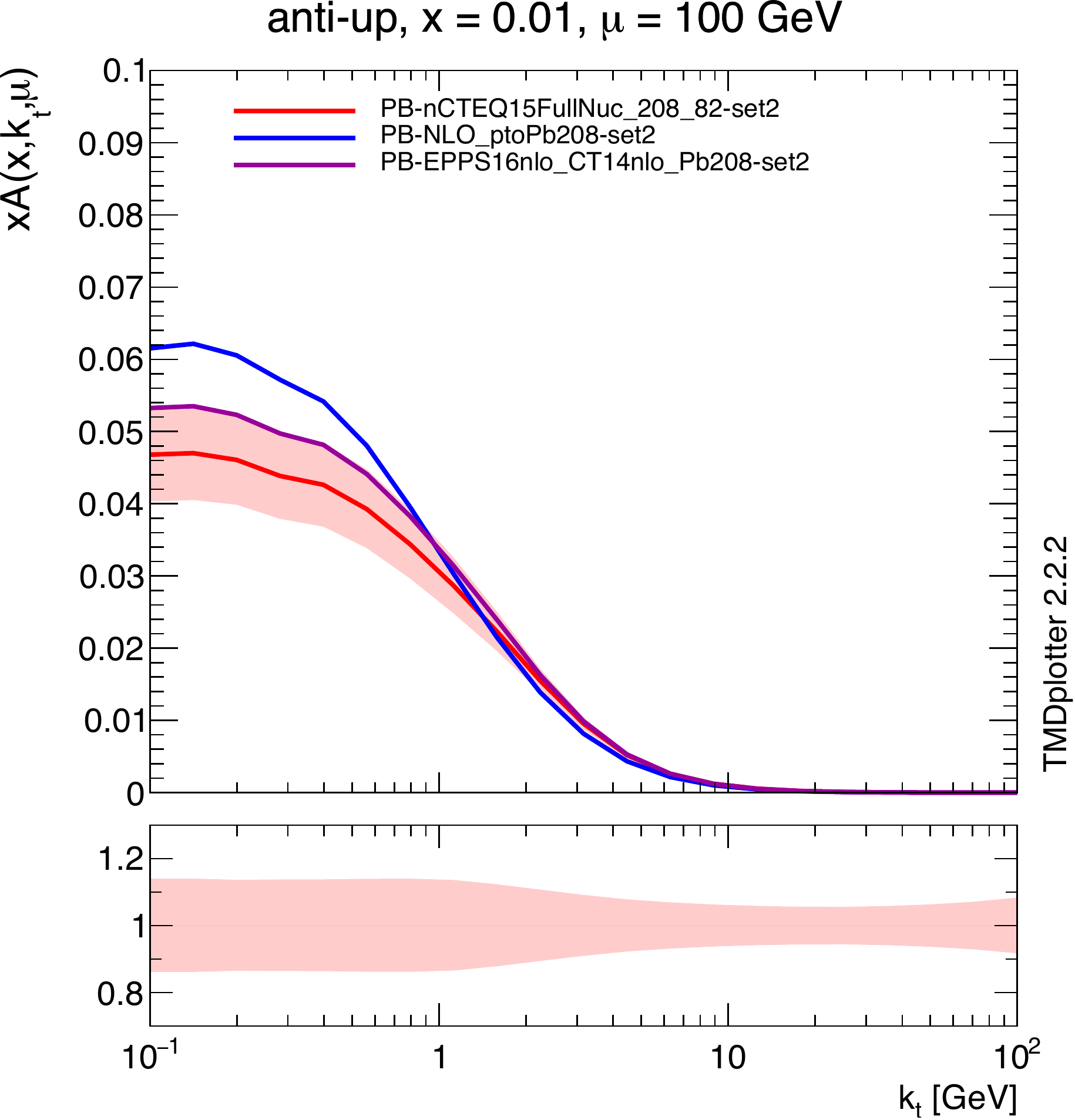}
\caption{\small
Transverse momentum dependent parton densities for $u$ and $\bar{u}$ quarks
at $x=0.01$ at the scale of $\mu=10$ GeV (upper plots) and $\mu=100$ GeV (lower plots).
The ratio shows relative uncertainty of \PBCTEQLead\ {\tt set2} TMDs.
}
\label{fig:TMD_pdfs_err}
\end{center}
\end{figure}

As a last part of this section, in Fig.~\ref{fig:TMD_pdfs_err}, we present
the uncertainties of the obtained nTMDs. We plot them for $u$ and $\bar{u}$
quarks at $x=0.01$ and two scales $\mu=10$~GeV and $\mu=100$~GeV. We can see that the
low $k_t$ region features the largest uncertainties: around 10\% for up and
15\% for $\bar{u}$ distributions. The size of the error band becomes smaller
with the evolution in $\mu$ but this reduction is still limited between the
two presented scales.
The distributions for other quarks exhibit similar uncertainties, on the other,
hand the uncertainties for the gluon are substantially larger reaching up to 50\%
at small $k_t$ values (covering also the differences between proton and nuclear
distributions).

\section{Results}
\label{sec:results}
We present now our predictions for the inclusive $Z$ boson production in $p$Pb
collisions at the LHC
\begin{equation}
p\mathrm{Pb} \to (Z/\gamma^*) \to \ell\bar{\ell}
\end{equation}
at $\sqrt{s}=5.02$ TeV and compare them with CMS
data~\cite{Khachatryan:2015pzs}. The intermediate vector boson is decaying into a pair of
electrons or muons and these two channels are combined and we compare with this
combined data. The measurement is done in the fiducial region defined by:
$p_T^\ell>20$ GeV, $|\eta^{\ell}_{\mathrm{lab}}|<2.4$ and $60<m_{\ell\ell}<120$ GeV.
In our calculations we use leading order (LO) off-shell matrix elements as calculated
by the Ka\hspace{-0.2ex}Tie Monte Carlo generator~\cite{vanHameren:2016kkz} and the TMDs (and PDFs)
discussed in Sec.~\ref{sec:pdfs}. The factorization and renormalization scales
are set to be equal to the $Z$ boson mass, $\mu=m_Z$.

The considered data is restricted to the $Z$ boson (center of mass) rapidity
ranging between $y^*\in(-2.8,2.0)$ which at LO corresponds to $x$ values in lead
$x_{\text{Pb}}\sim(3*10^{-3},0.3)$ (where negative rapidities correspond to
large $x_{\text{Pb}}$ and positive $y^*$ correspond to small $x_{\text{Pb}}$).
This shows that the probed $x$ values are not very small and we do not need
to worry about non-linear effects.

\begin{figure*}[!t]
\centering{}
\subfloat[]{
\includegraphics[width=0.49\textwidth]{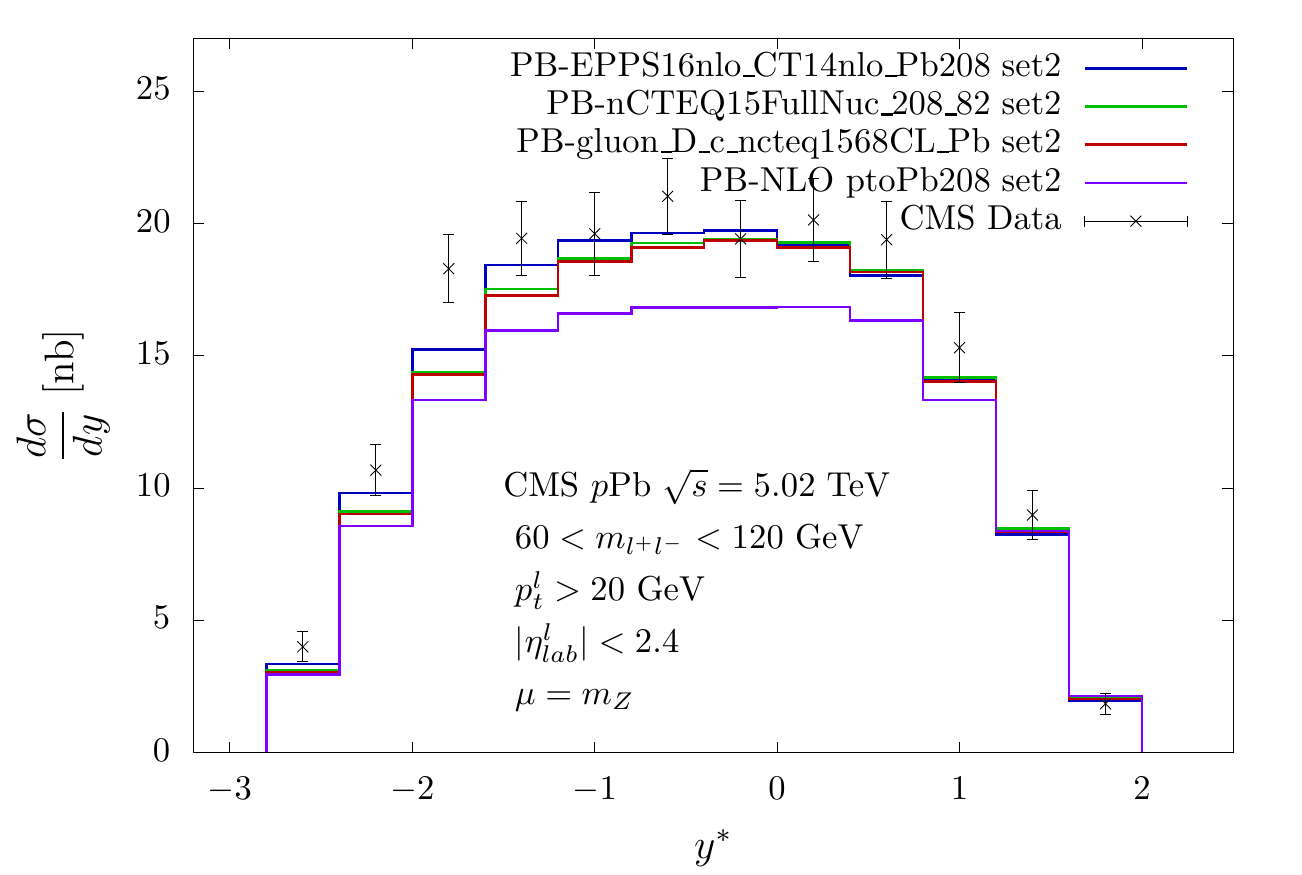}}
\hfil
\subfloat[]{
\includegraphics[width=0.49\textwidth]{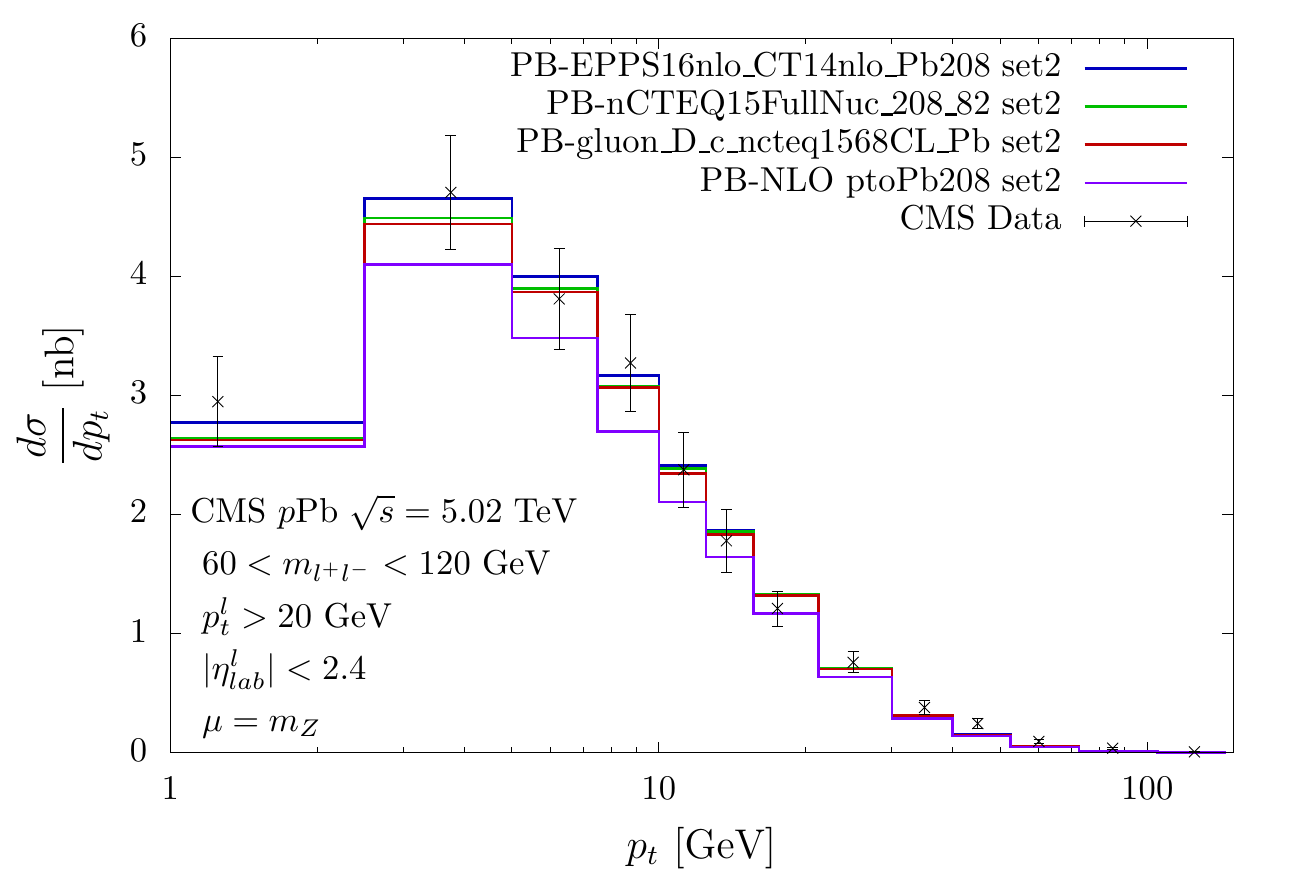}}
\caption{Comparison of predictions for:
(a) $Z$ boson (center of mass) rapidity $y^*$, and
(b) $Z$ boson transverse momentum, $p_T$,
distributions obtained within $k_T$-factorization using different PB nTMDs/TMDs
with the CMS data~\cite{Khachatryan:2015pzs}. }
\label{fig:compare_set2}
\end{figure*}

In Fig.~\ref{fig:compare_set2}
we first present a comparison of predictions obtained
using different PB TMDs for lead and proton~\cite{Hautmann:2017fcj}.
In all cases we use  {\tt Set2} distributions. We can see a very good description of the
data provided by the $k_T$-factorization framework. 
This is true for all the nuclear TMDs (\PBCTEQLead, \PBEPPSLead, PB-gluon\_D\_c\_ncteq1568CL\_Pb).
We observe only minor differences between predictions obtained with different
nTMDs which suggest that the details of the nuclear corrections are here of
secondary importance
compared to the framework itself. On the other hand, we can see that neglecting the
nuclear correction entirely (using lead composed out of free-proton TMDs -- \PBproton)
undershoots the measured cross sections as a function  of the rapidity and the transverse momentum.
One should highlight that these are absolute distributions meaning that the LO
framework we are using predicts not only the shape of the distributions but also
their normalization (the uncertainty of the prediction is discussed at the end of this section).
%

\begin{figure*}[!t]
\centering{}
\subfloat[]{
\includegraphics[width=0.49\textwidth]{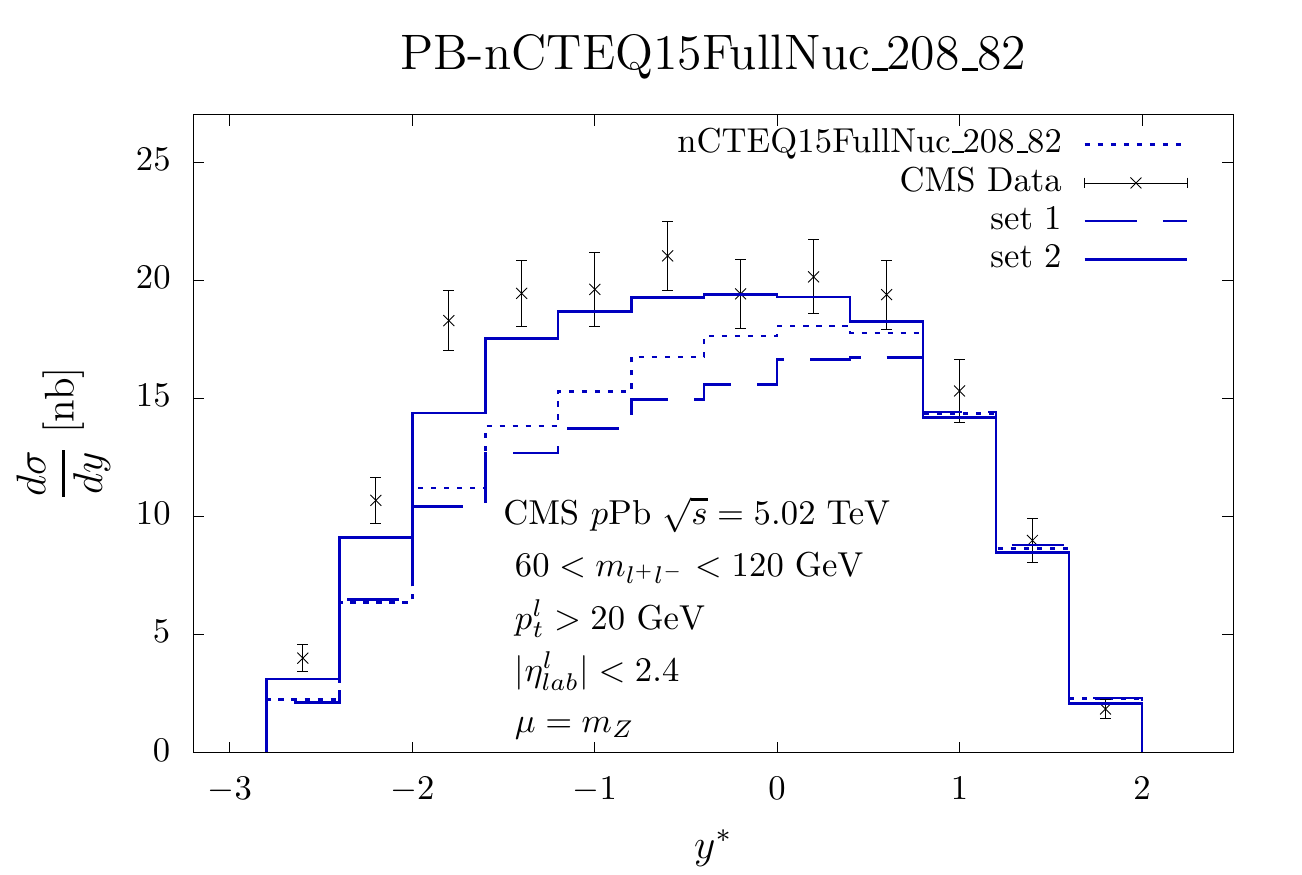}}
\hfil
\subfloat[]{
\includegraphics[width=0.49\textwidth]{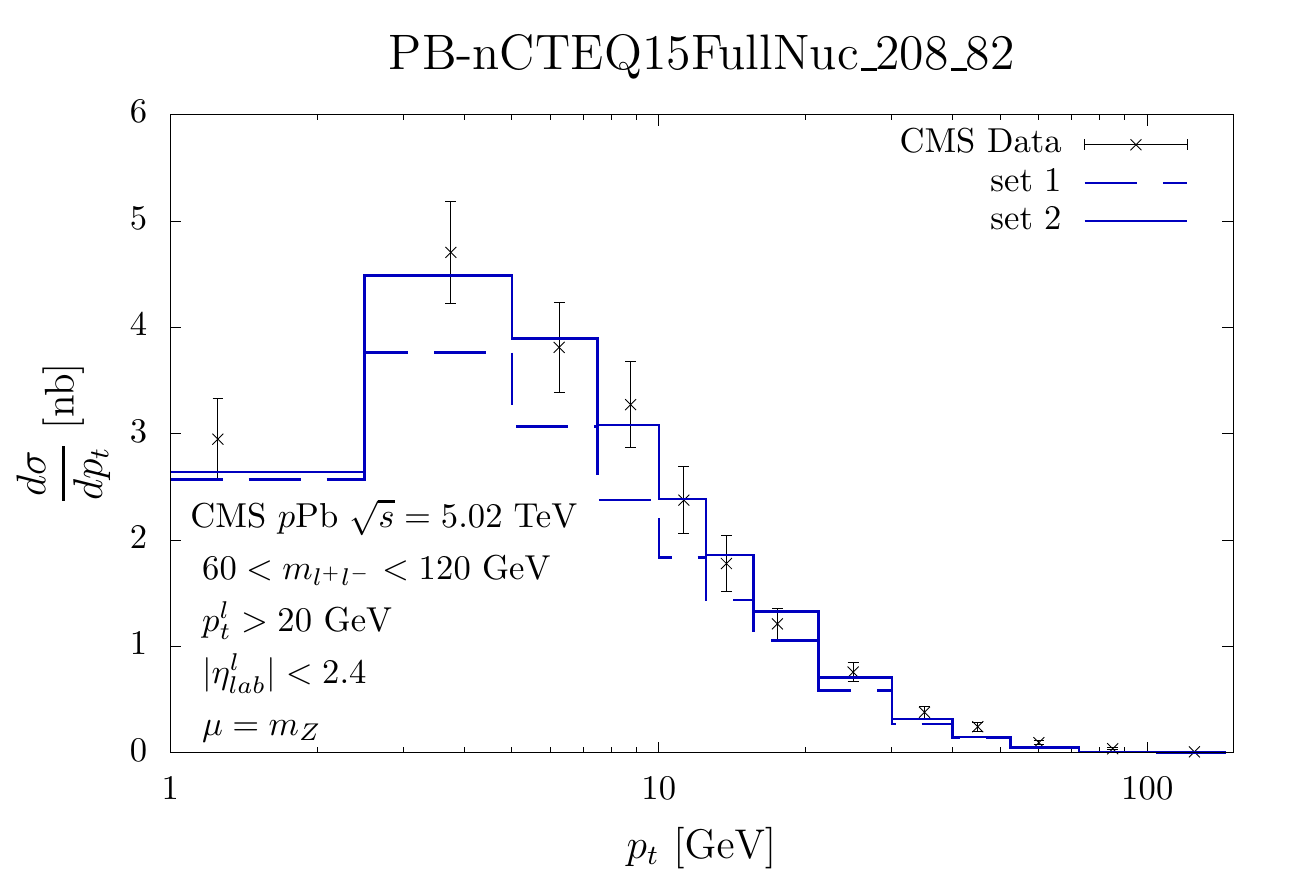}}
\caption{Comparison of predictions for:
(a) $Z$ boson (center of mass) rapidity $y^*$, and
(b) $Z$ boson transverse momentum, $p_T$,
distributions obtained within $k_T$-factorization using \PBCTEQLead\ {\tt Set1} and {\tt Set2} nTMDs
and within collinear factorization using nCTEQ15 nPDFs (nCTEQ15FullNuc\_208\_82)
with the CMS data~\cite{Khachatryan:2015pzs}.}
\label{fig:ncteq15_set1-2}
\end{figure*}

In order to understand the obtained results in more detail we perform an additional
comparison of our predictions with LO collinear results as well as with the hybrid
approach~\cite{Dumitru:2005gt, Deak:2009xt,Kotko:2015ura}.%
    \footnote{Strictly speaking the hybrid approach is valid in the situation
    of more exclusive observables where one of the partons involved in the scattering
    has low-$x$ and the other not, e.g.\ for jet production in the forward region.
    Here it will serve us rather as a tool for better understanding the
    $k_T$-factorization results.}
The results are very similar for all the nuclear TMDs and therefore we concentrate
only on the ones obtained with the \PBCTEQLead\ distributions.
In Fig.~\ref{fig:ncteq15_set1-2} we first compare the $k_T$-factorization results
obtained with the \PBCTEQLead\ TMDs of {\tt Set1} and {\tt Set2} with the corresponding collinear
predictions (of course, in the collinear case at LO the $p_T$ of the $Z$ boson vanishes).
We clearly see that the rapidity distribution obtained in the collinear - and
$k_T$-factorization with {\tt Set1} TMDs are very similar but give much worse
description of the measurement compared to the calculation performed in $k_T$-factorization calculation with {\tt Set2}.
The similarity of the collinear and TMD {\tt Set1} results is not surprising as the {\tt Set1} TMDs
reduce to the collinear distributions after the integration over the transverse momentum,
and for the rapidity distributions such an integration is effectively carried out 
(with additional corrections coming from the off-shell matrix elements). 
Interestingly, the results obtained with {\tt Set2} are very different and describe the data much
better than the {\tt Set1}. The same holds for the $p_T$ distribution where results obtained with
{\tt Set2} is spot on the data whereas the {\tt Set1} results are lower.
%

\begin{figure*}[!t]
\centering{}
\subfloat[]{
\label{fig:ncteq15_hybrid2_rap}
\includegraphics[width=0.49\textwidth]{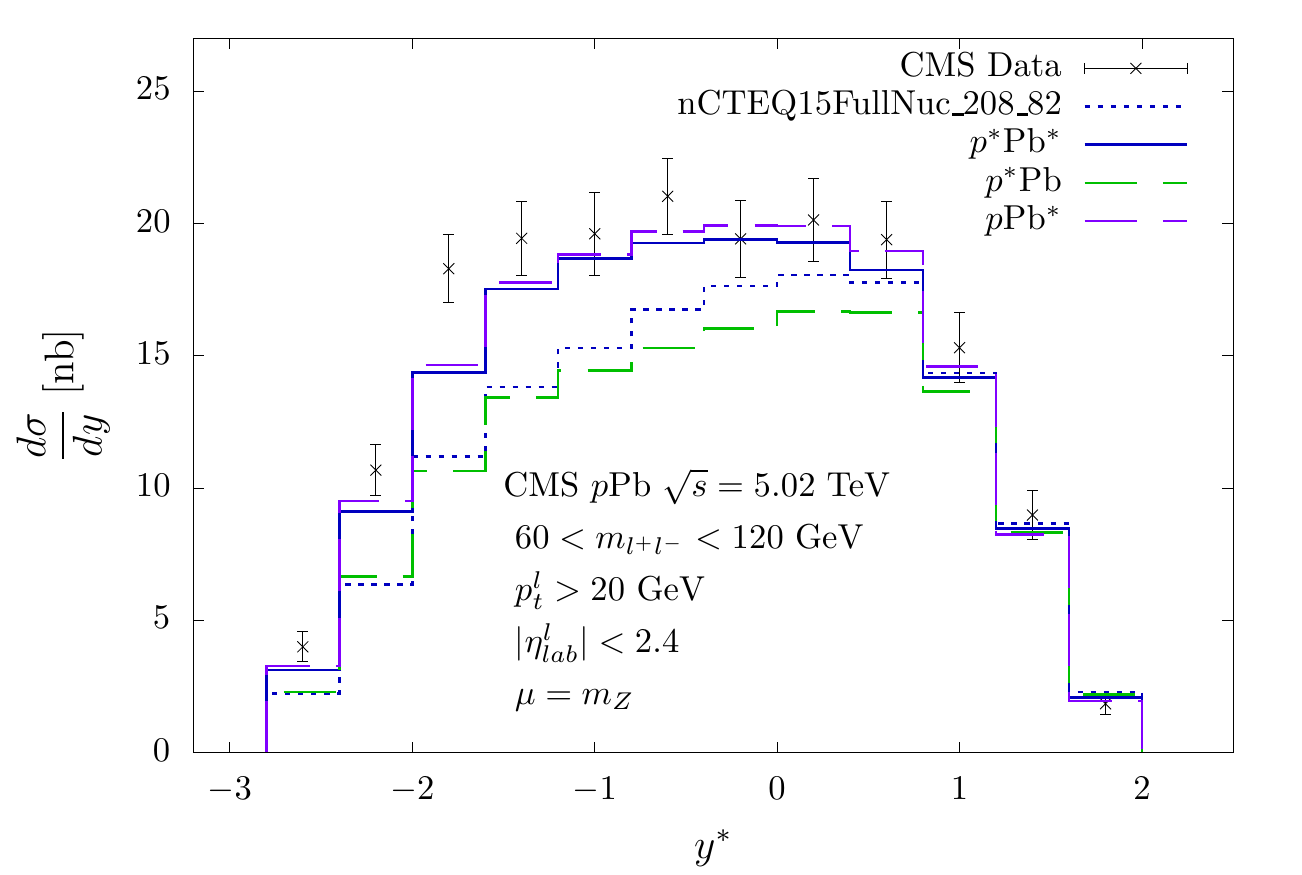}}
\hfil
\subfloat[]{
\label{fig:ncteq15_hybrid2_pt}
\includegraphics[width=0.49\textwidth]{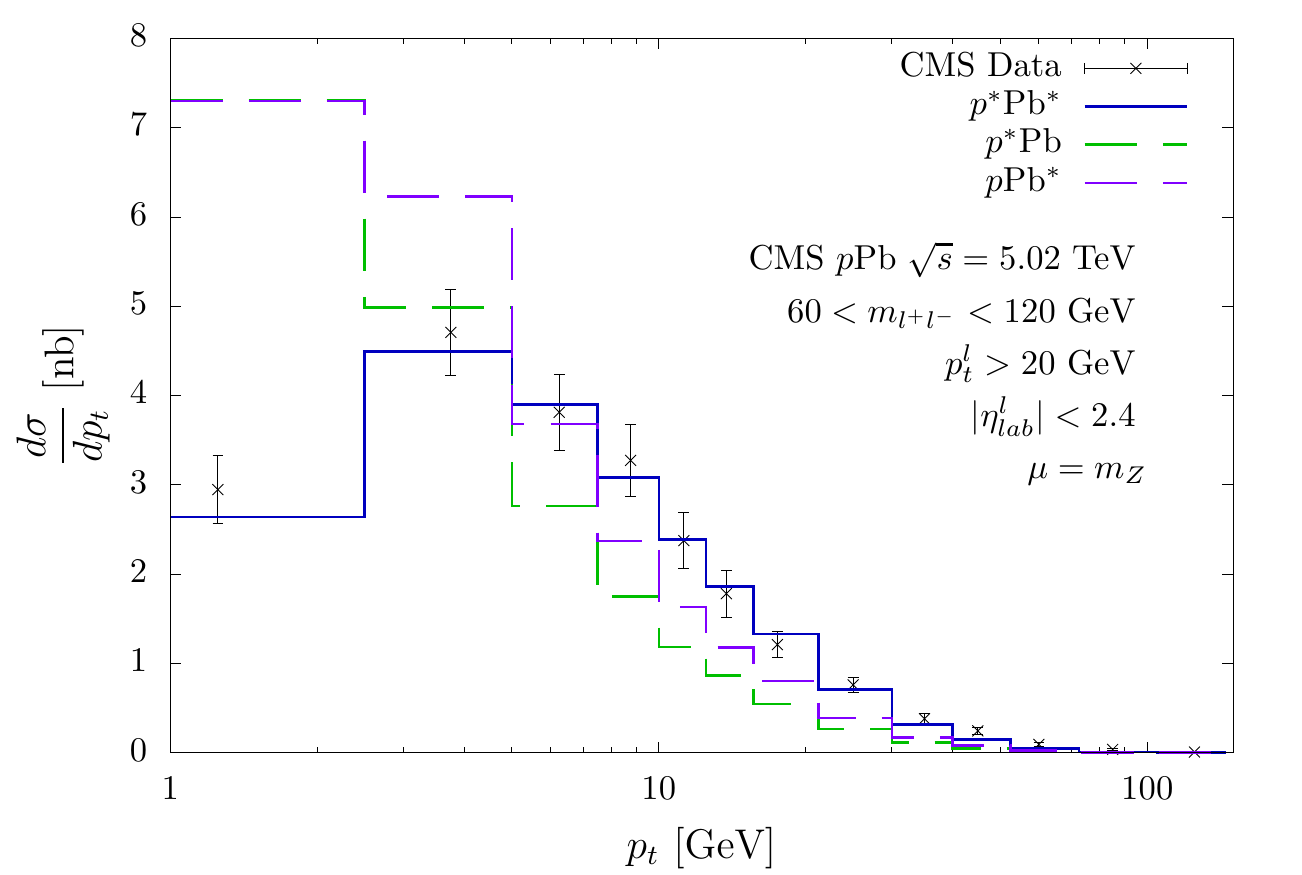}}
\caption{Comparison of predictions for:
(a) $Z$ boson (center of mass) rapidity $y^*$, and
(b) $Z$ boson transverse momentum, $p_T$,
distributions obtained using nCTEQ15 TMDs/PDFs with $k_T$-factorization (TMD),
hybrid approach with two choices for the off-shell initial partons ($p^*$Pb, $p$Pb$^*$)
and collinear approach.
The results are compared with the CMS data~\cite{Khachatryan:2015pzs}.}
\label{fig:ncteq15_hybrid2}
\end{figure*}

To further analyze the situation in Fig.~\ref{fig:ncteq15_hybrid2} we do an additional
comparison using only {\tt Set2} nTMDs but in addition to the $k_T$-factorization and collinear
results we include also results from the hybrid approach. In the hybrid approach the parton
distributions of one beam are given by regular collinear PDFs whereas the second beam
is described by the TMDs. Since we are in the central region
($|\eta^{\ell}_{\mathrm{lab}}|<2.4$), a priori there is no clear choice on which beam
should be described by the TMD and which by the collinear PDF. We present
computations for both choices.
Rather surprisingly we can see from the rapidity distribution in Fig.~\ref{fig:ncteq15_hybrid2_rap}
that the hybrid calculation with an off-shell parton from the  lead beam using the nTMDs gives
results very close to the calculation  with both partons being off shell (as in the $k_T$-factorization calculation). 
On the other hand
the hybrid calculation with an off-shell parton from the proton beam gives results very close to the collinear
calculation. In the $p_T$ distribution, the hybrid result with an off-shell parton from the lead
beam is also closer to the data, however, at low $p_T$ both hybrid results fail to describe
the Sudakov suppression and fail to describe the data (which are very well described when both 
partons are treated off-shell in the calculation in $k_T$-factorization).
%

\begin{figure*}[!t]
\centering{}
\includegraphics[width=0.49\textwidth]{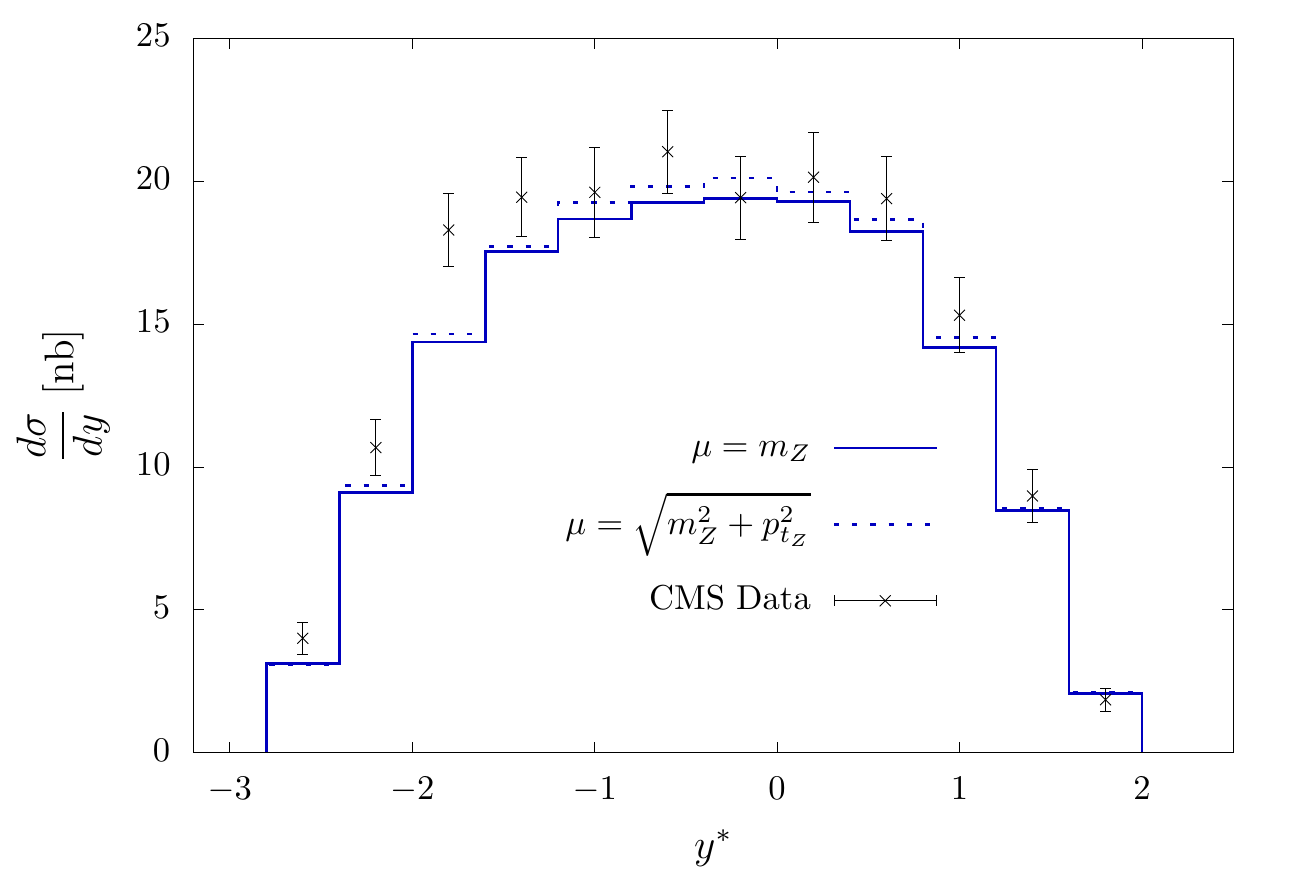}
\includegraphics[width=0.49\textwidth]{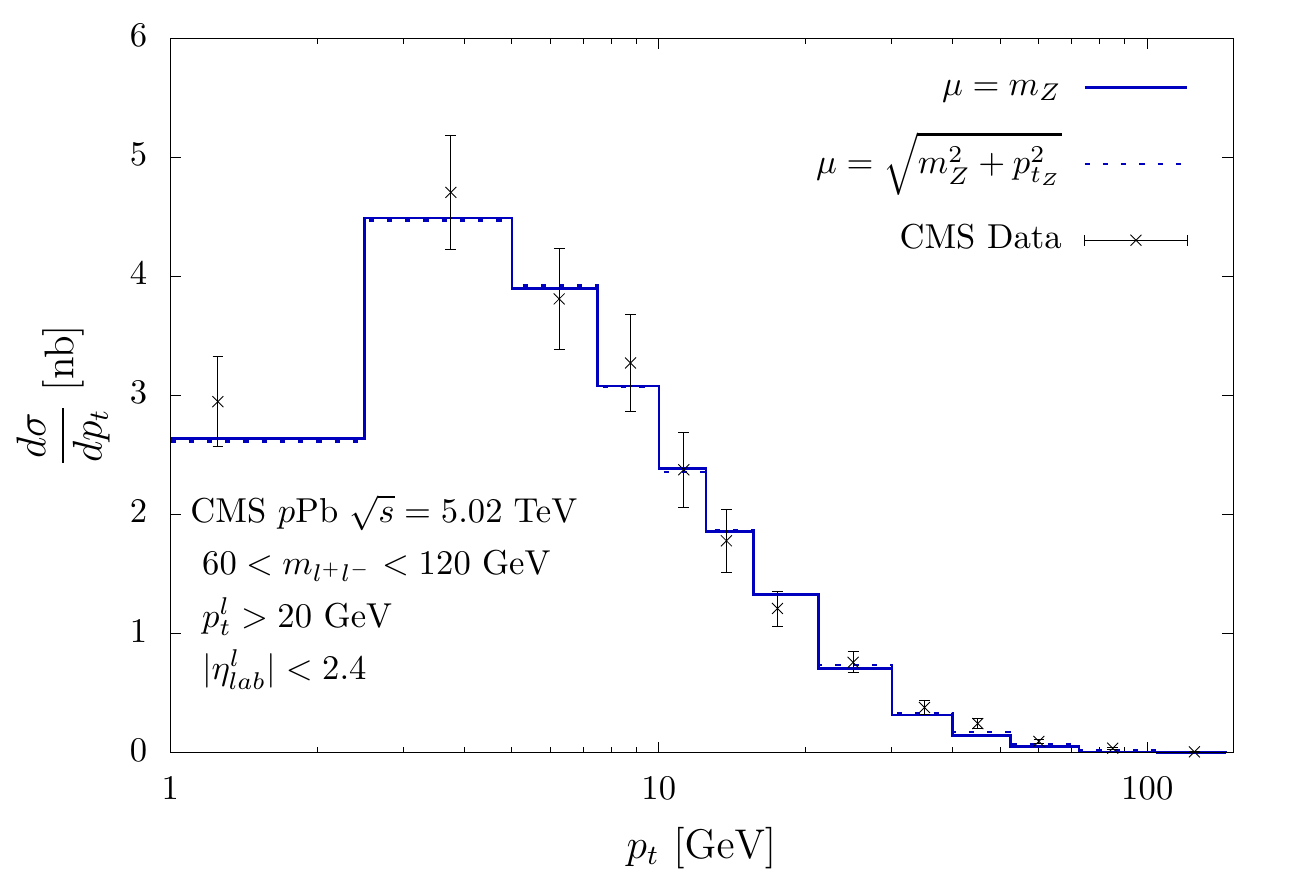}
\caption{Comparison of $Z$ boson rapidity and transverse momentum distributions
computed with scale $\mu=m_Z$ and $\mu=\sqrt{m_Z^2+p_{tZ}^2}$ using \PBCTEQLead\ {\tt Set2} TMDs.}
\label{fig:scaleComp}
\end{figure*}

\begin{figure*}[!htb]
\centering{}
\includegraphics[width=0.49\textwidth]{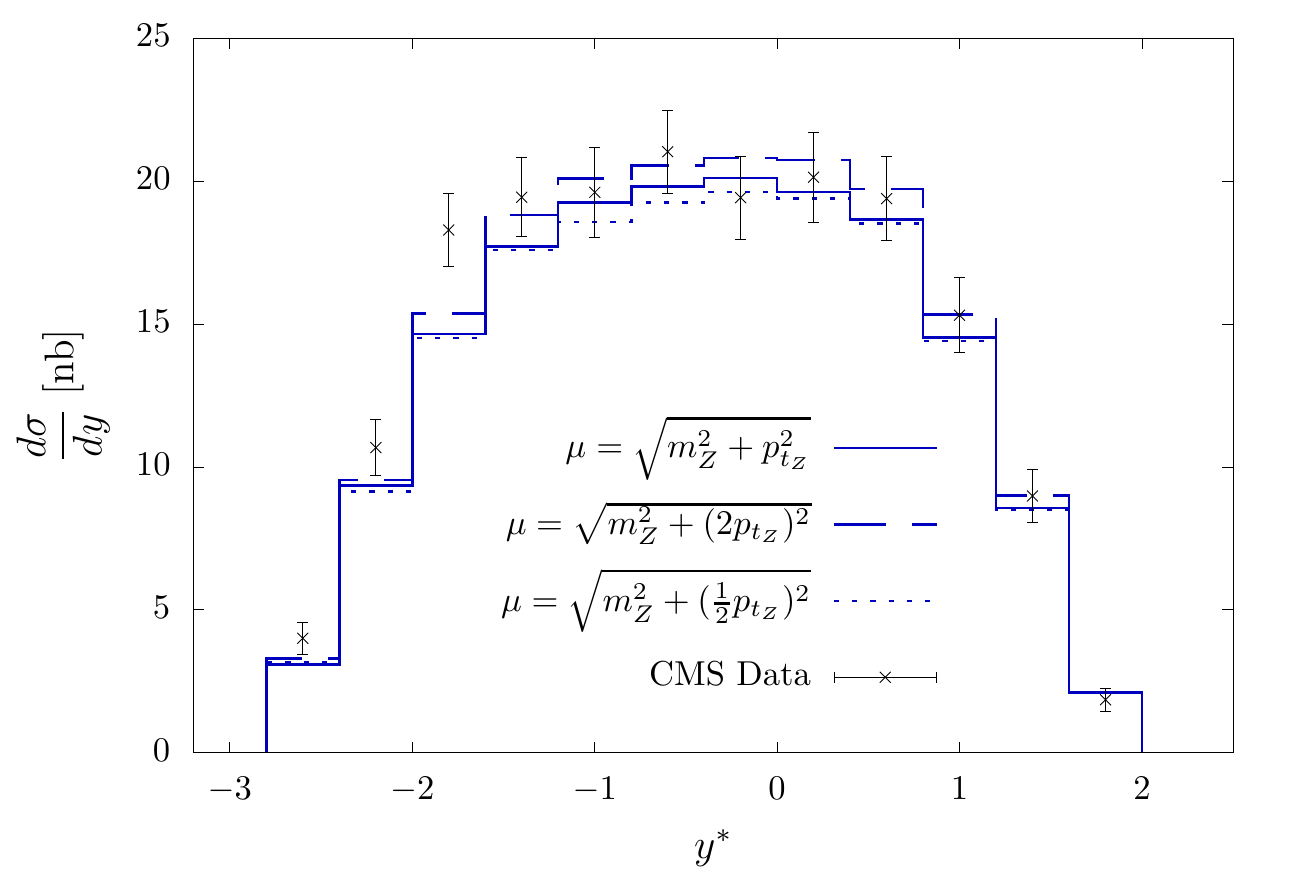}
\includegraphics[width=0.49\textwidth]{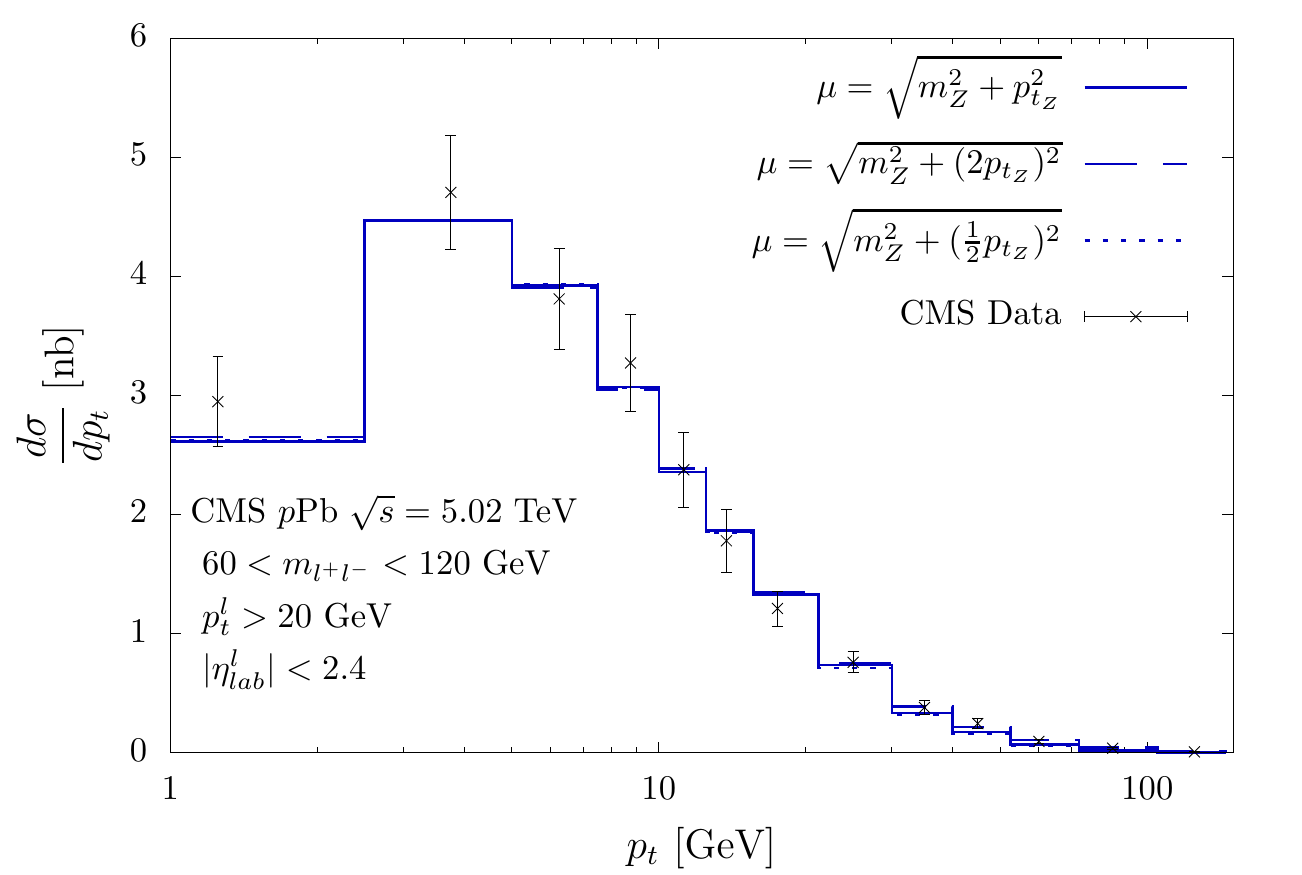}
\caption{Scale variation for the predictions for $Z$ boson rapidity and transverse
momentum distributions compute with \PBCTEQLead\ {\tt Set2} TMDs.}
\label{fig:scaleVar}
\end{figure*}

In order to estimate missing higher orders it is conventional to perform
a scale variation by a factor two up and down. 
Higher order contributions come from the radiation of additional partons, and therefore will
depend on the transverse momentum of the \PZ\-boson: at $p_T=0$ no real parton can be emitted
and the scale variation should lead to minimal changes. However, at larger $p_T$ the scale variation
should give significant effects. We therefore introduce a scale $\mu^2 = m_{\PZ }^2 + p_T^2$,
where we vary only the transverse momentum in order to estimate the effects from missing higher orders.
A similar choice of the scale and scale variation was applied in Ref. \cite{Dooling:2014kia}.
In Fig.~\ref{fig:scaleComp} we show a comparison of calculations using two different choices for the
factorization scales $\mu^2 = m_{\PZ }^2$ (that we used in earlier plots) and $\mu^2 = m_{\PZ }^2 + p_T^2$; the predictions are
very similar. In Fig.~\ref{fig:scaleVar} we show the result of the scale variation with
$\mu^2 = m_{\PZ }^2 + p_T^2$,
where the dynamical part ($p_T$) is varied by a factor of two up and down. The scale dependence is visible
in the rapidity distribution and  at larger transverse momenta (as expected). The size of the variation is
similar to what is expected in NLO calculations.
%
%

\begin{figure*}[!htb]
\centering{}
\includegraphics[width=0.49\textwidth]{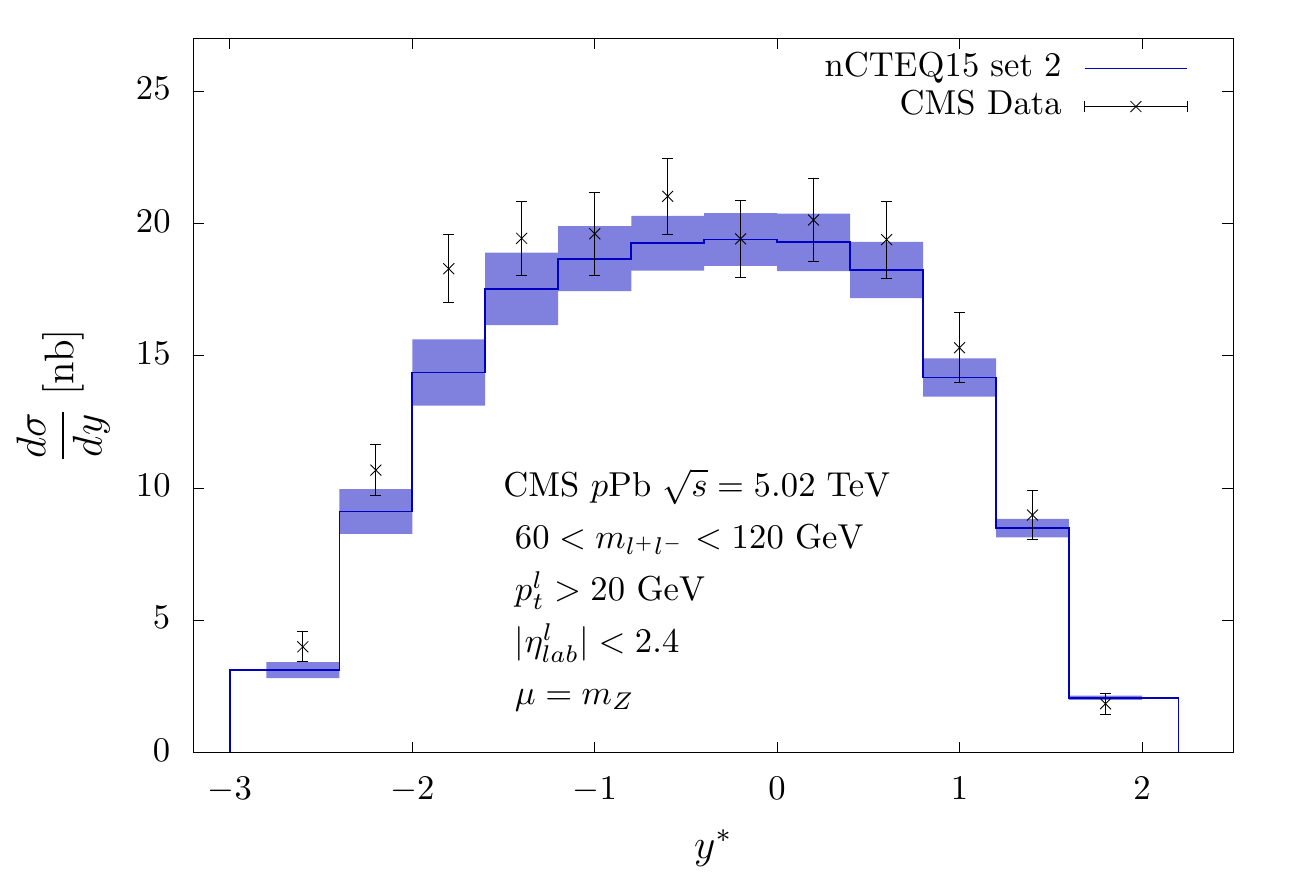}
\includegraphics[width=0.49\textwidth]{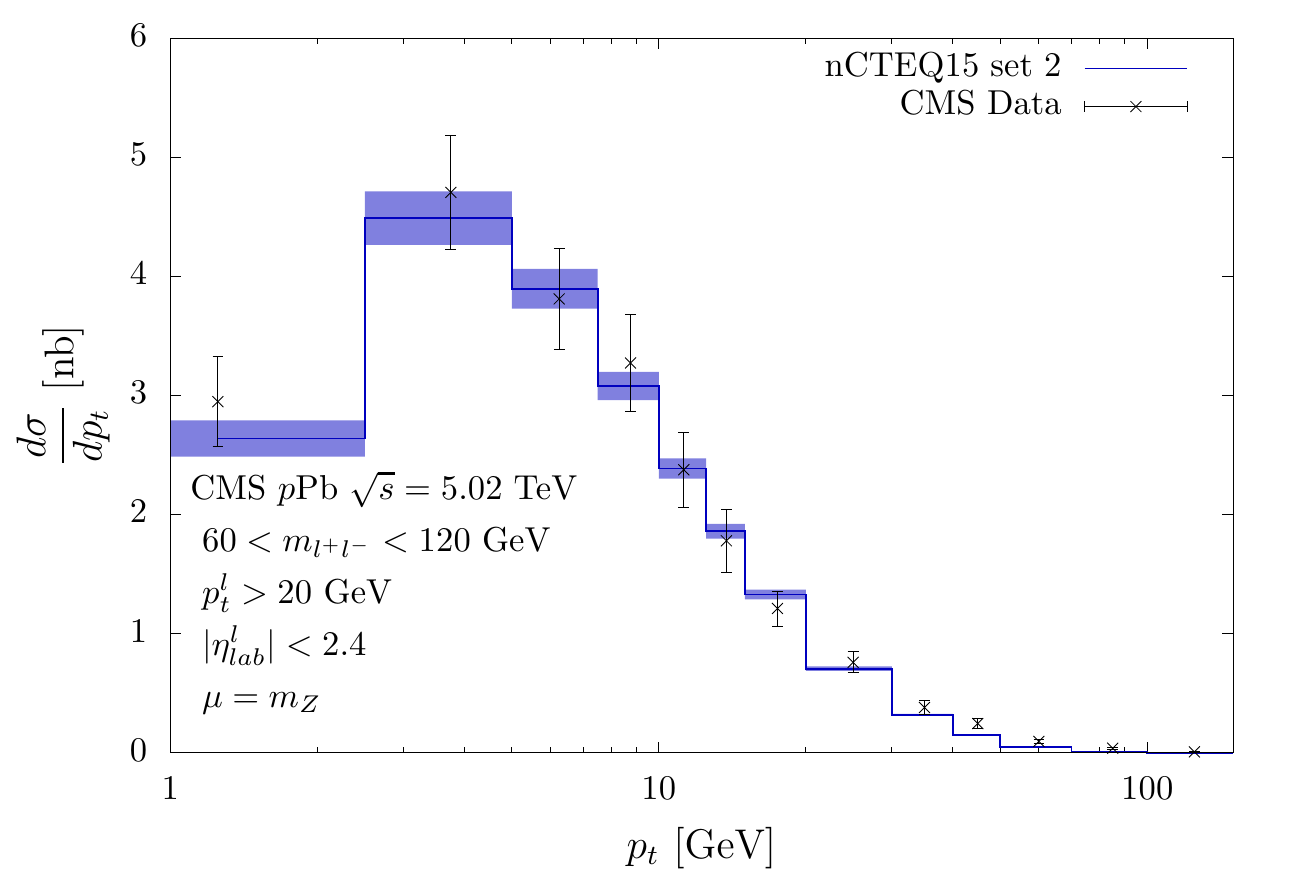}
\caption{
TMD uncertainties for the predictions for $Z$ boson rapidity and transverse
momentum distributions compute with \PBCTEQLead\ {\tt Set2} TMDs.}
\label{fig:pdfErr}
\end{figure*}

Finally, in Fig~\ref{fig:pdfErr} we show the effect of the \PBCTEQLead\
{\tt Set2} TMD uncertainties on the rapidity and transverse momentum distributions
of the $Z$ boson. One can see that the TMD uncertainties are larger than the scale
variation. They are also larger than the spread of predictions using different
nTMDs (shown in Fig.~\ref{fig:compare_set2}) but not wide enough to cover also
the predictions with proton TMDs (also in Fig.~\ref{fig:compare_set2}). This
again highlights that the use of nuclear corrections is important but with the
current experimental errors their details are secondary.
%


\subsection{Predictions for 8 TeV data}
\begin{figure*}[!t]
\centering{}
\subfloat[$60<m_{ll}<120$ GeV]{
\label{fig:pred8tev_mZ}
\includegraphics[width=0.49\textwidth]{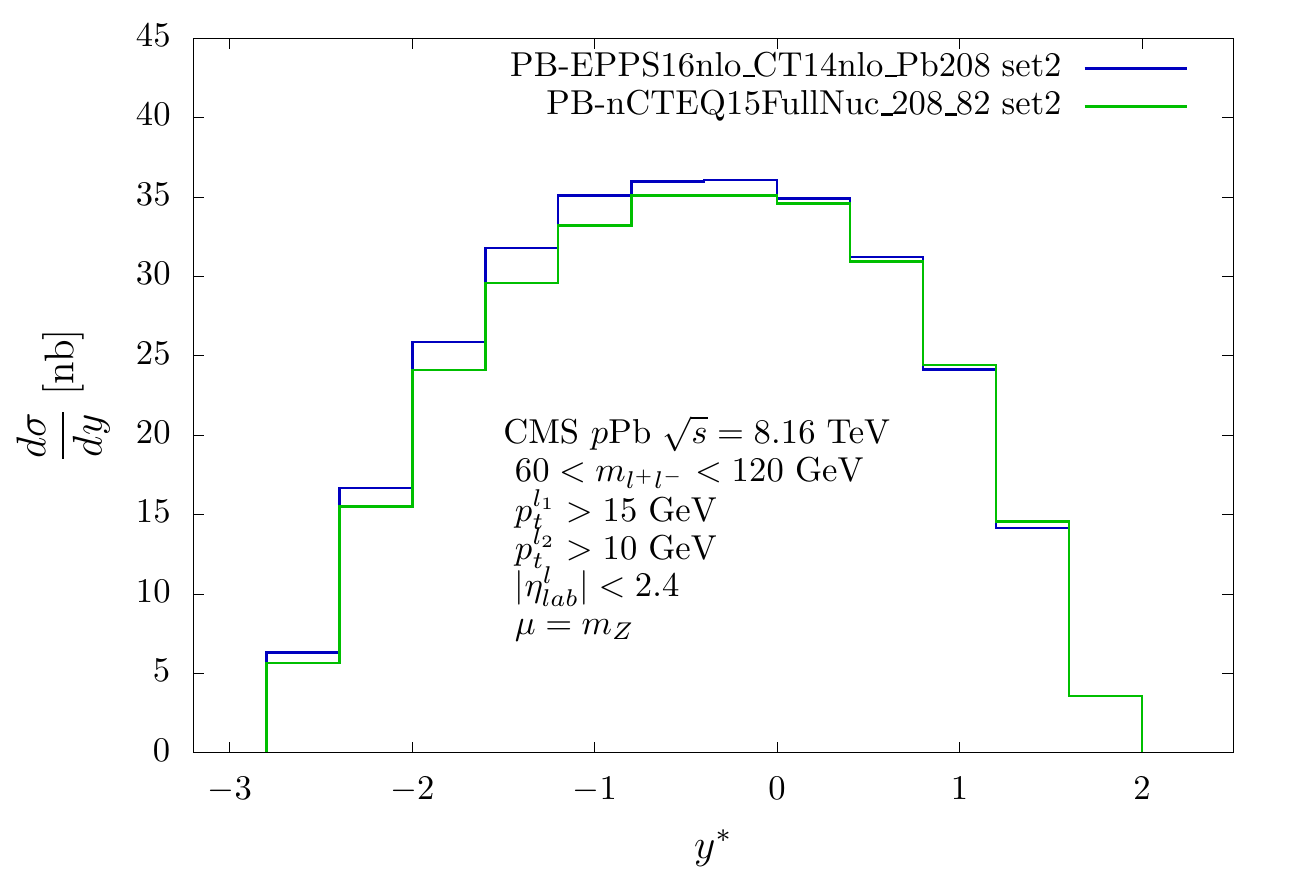}
\includegraphics[width=0.49\textwidth]{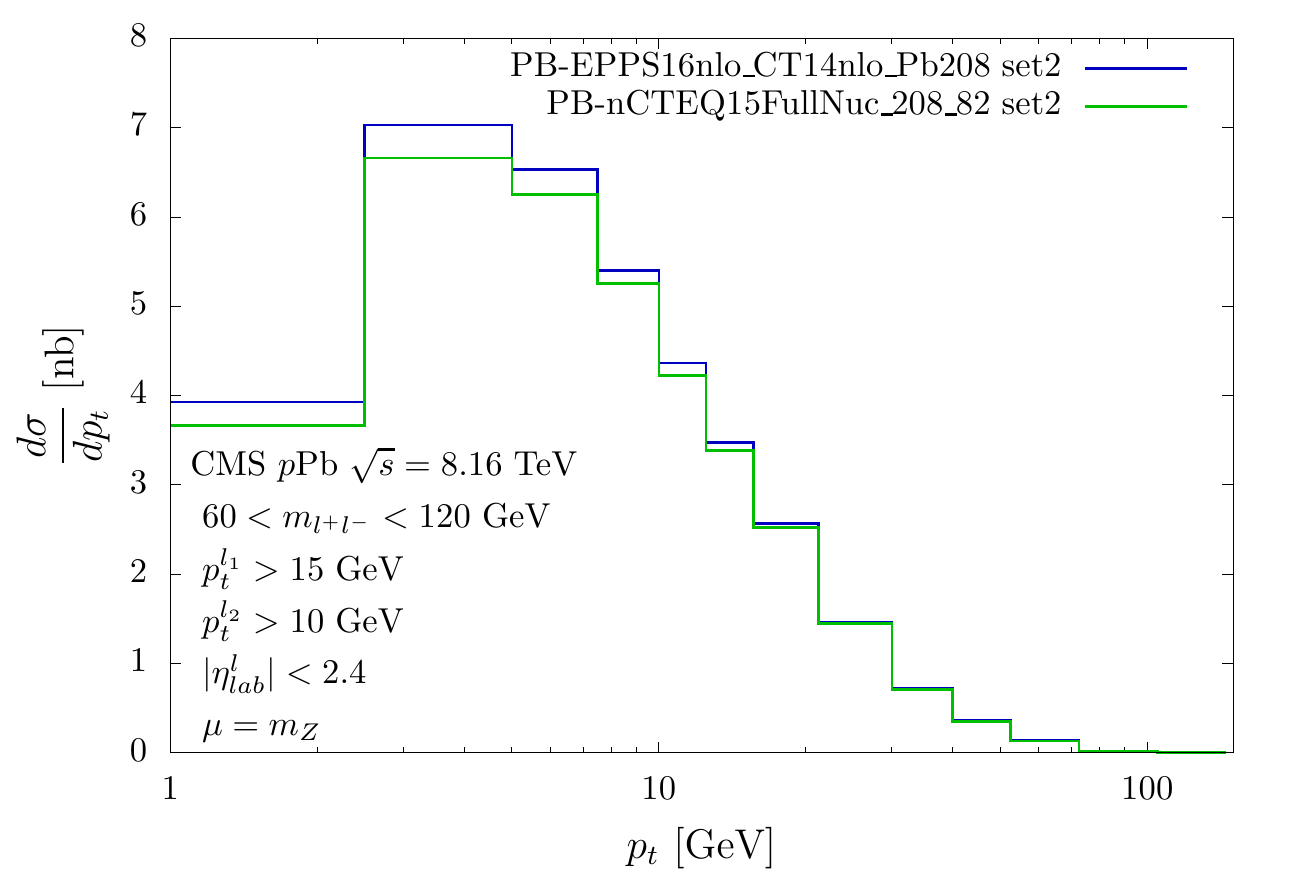}}
\hfil
\subfloat[$15<m_{ll}<60$ GeV]{
\label{fig:pred8tev_fullRange}
\includegraphics[width=0.49\textwidth]{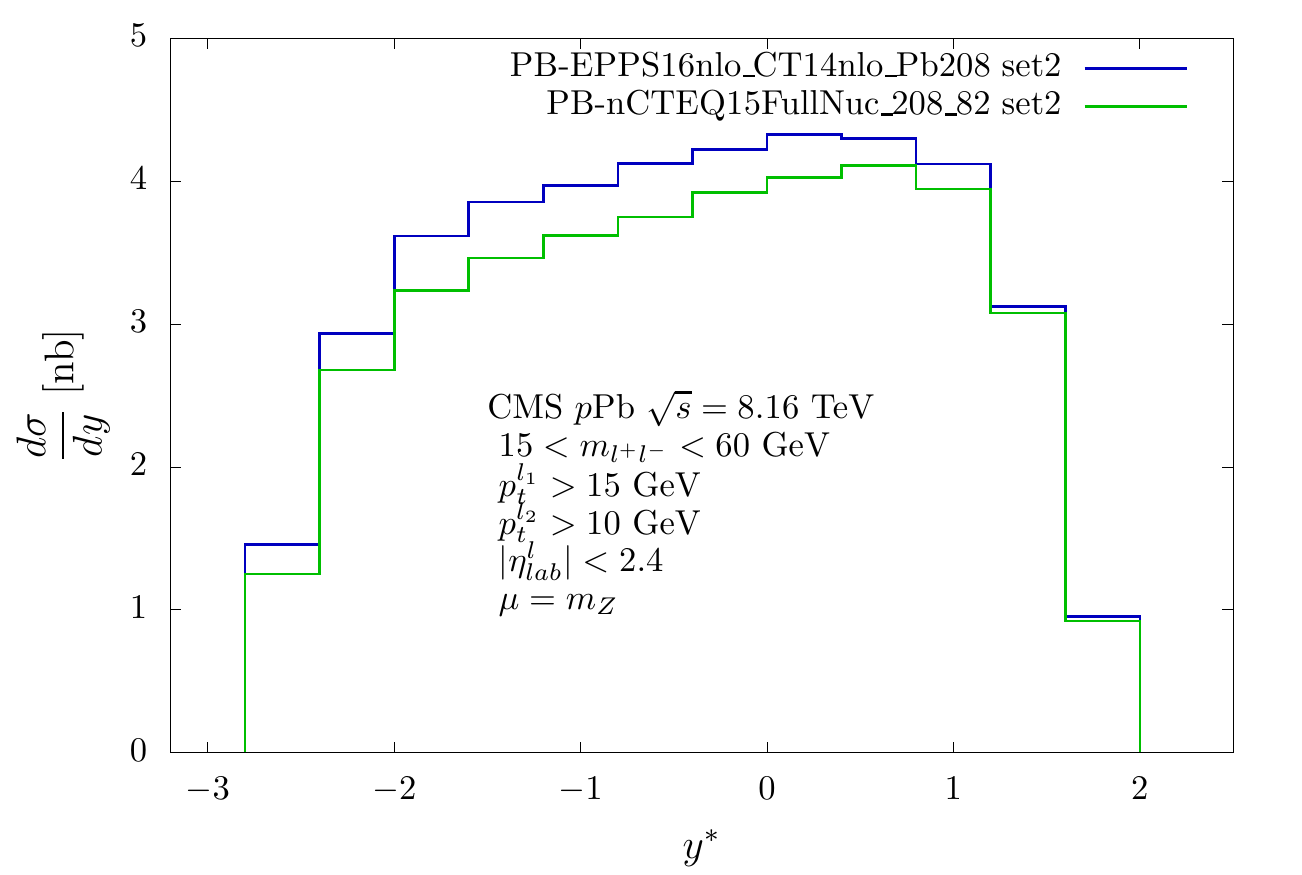}
\includegraphics[width=0.49\textwidth]{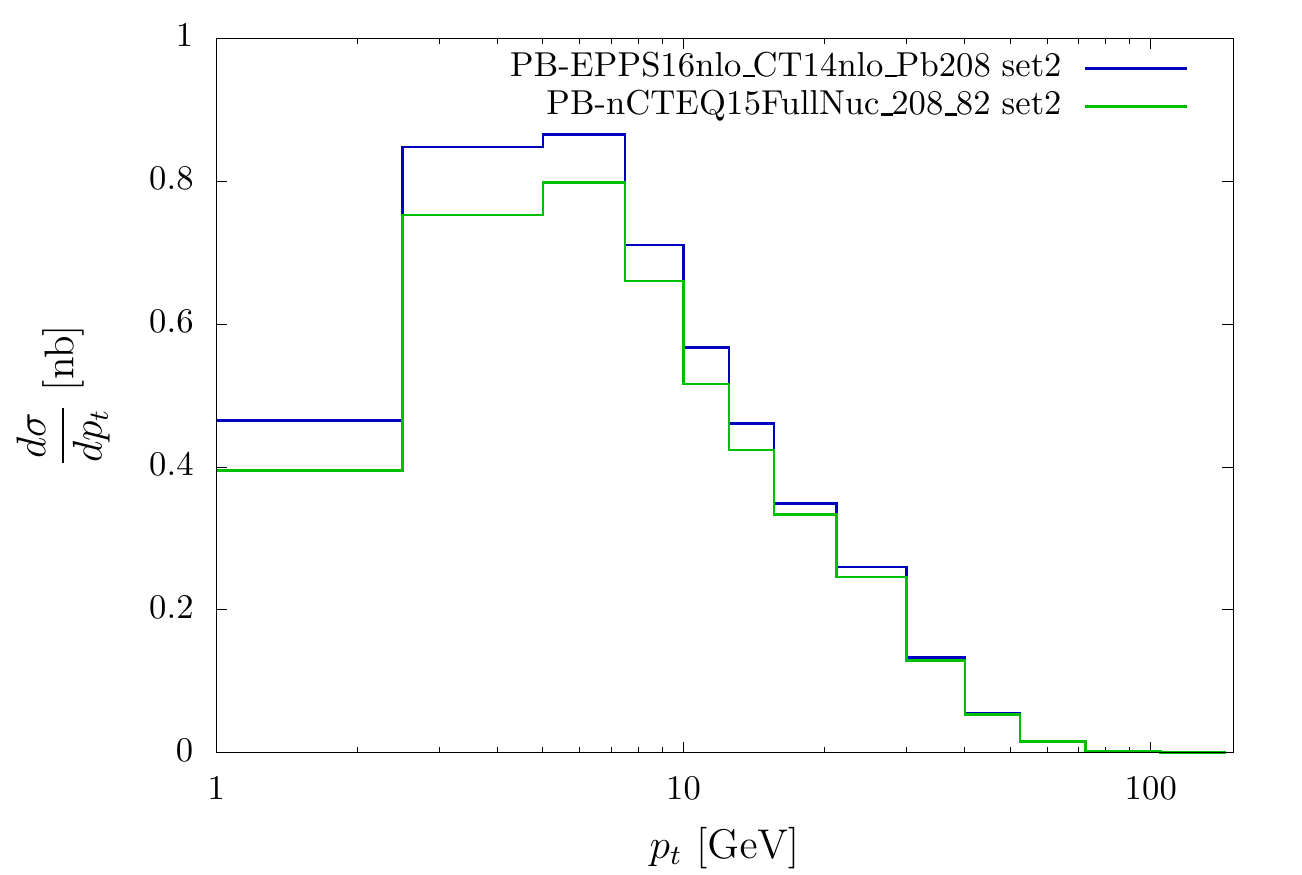}}
\caption{
Predictions for $Z$ boson rapidity and transverse
momentum distributions in $p$Pb collisions at $\sqrt{s}=8.16$ TeV
in the invariant lepton pair mass range
(a) $60<m_{ll}<120$ GeV,
and (b) $15<m_{ll}<60$ GeV
computed with \PBCTEQLead\ and \PBEPPSLead\ {\tt Set2} TMDs.}
\label{fig:pred8tev}
\end{figure*}

The data for $Z$ boson production in proton-lead collisions at 8.16 TeV are
currently being analyzed by LHC collaborations. As an addition to the results
presented above we also provide prediction for such measurements. 
In Fig.~\ref{fig:pred8tev} we show both
rapidity and transverse momentum distributions of the produced $Z$ boson
(lepton pair). We present results for two ranges of the invariant mass of the
produced lepton pair: around the $Z$ boson mass peak $60<m_{ll}<120$ GeV,
and for low mass DY pair $15<m_{ll}<60$ GeV.

%

\section{Conclusion}
\label{sec:conclusions}
We have constructed the first sets of nuclear TMDs for gluons as well as all
the quarks. It was done using the parton branching method employing different
choices of $\alpha_S$ arguments. The obtained nTMDs are provided to all interested
colleagues on the TMDlib~\cite{Hautmann:2014kza} website \url{https://tmdlib.hepforge.org}.
We have used the constructed nTMDs in the $k_T$-factorization framework to calculate
predictions for $Z$ boson production at $p$Pb collisions at the LHC. The obtained
results show very good description of the CMS data~\cite{Khachatryan:2015pzs} for
both rapidity and $p_T$ distributions. One should highlight here that predictions
are not only for the shape of the distributions but also for their normalization.

\section*{Acknowledgments}
 Etienne Blanco and Aleksander Kusina acknowledges the partial support of Narodowe Centrum Nauki with grant DEC-2017/27/B/ST2/01985 Krzysztof Kutak and Andreas van Hameren acknowledge the partial support by COST Action CA16201 {\em Unraveling new physics at the LHC through the precision frontier}.

\providecommand{\href}[2]{#2}\begingroup\raggedright\endgroup


\end{document}